\documentclass[11pt]{article}
\usepackage{graphicx}

\def \n{\hfil\break}

\def \Nbf{{\bf N}}

\def \qh{{\hat q}}
\def \be{\begin{equation}}
\def \ee{\end{equation}}
\def \d{\partial}
\def \aa{\alpha}
\def \bb{\beta}
\def \dd{{\delta}}
\def \gg{\gamma}
\def \ss{\sigma}
\def \ll{\lambda}

\def \eps{\epsilon}
\def \kk{\kappa}
\def \DD{\Delta}
\def \SS{\Sigma}
\def \LL{\Lambda}
\def \GG{\Gamma}
\def \Om{\Omega}

\def \Ab{{\bar A}}

\def \Fc{{\cal F}}
\def \Mc{{\cal M}}
\def \Mch{{\hat \Mc}}

\def \Fc{{\cal F}}

\def \Vc{{\cal V}}
\def \Mcb{{\bf \Mc}}

\def \fr{\frac}

\def \veps{{\varepsilon}}

\def \cb{{\bar c}}

\def \gb{{\bar g}}
\def \ggb{{\bar \gg}}
\def \kkb{{\bar \kk}}
\def \ght{{\hat g}}

\def \eb{{\bar e}}

\def \mb{{\bar m}}
\def \aab{{\bar \aa}}
\def \bbb{{\bar \bb}}
\def \mub{{\bar \mu}}
\def \llb{{\bar \ll}}
\def \mub{{\bar \mu}}

\def \Ab{{\bar A}}

\def \Pb{{\bar P}}

\def \psib{{\bar \psi}}
\def \etab{{\bar \eta}}
\def \Tr{{\hbox{Tr}}}
\def \Ut{{\tilde{U}}}
\begin{document}
\title{Bimetric QED }
\author{I.T. Drummond\thanks{email: itd@damtp.cam.ac.uk} \\
            Department of Applied Mathematics and Theoretical Physics\\
            Centre for Mathematical Sciences\\
            Wilberforce Road\\ Cambridge\\ England, CB3 0WA
        }
\maketitle

\abstract{
We study, as a model of Lorentz symmetry breaking, the quantisation and renormalisation of
an extension of QED in a flat spacetime where the photons and electrons propagate 
differently and do not share the same lightcone. We will refer to this model as 
Bimetric QED (BIMQED). As a preliminary we discuss the formulation 
of electrodynamics in a pre-metric formalism showing nevertheless that there 
is, on the basis of a simple criteron, a {\it preferred} metric. Arising from this choice of metric 
is a Weyl-like tensor (WLT). The Petrov classification of the WLT gives rise to a 
corresponding classification of Lorentz symmetry breaking. We do not impose any constraint on the
strength of the symmetry breaking and are able to obtain explicit dispersion relations for
photon propagation in each of the Petrov classes. The associated birefringence appears in
some cases as two distinct polarisation dependent lightcones and in other cases as a
a more complicated structure that cannot be disentangled in a simple way. 

We show how in BIMQED the renormalisation procedure can, in addition to its effect on standard 
parameters such as charge and mass, force the renormalisation of the metrics and the WLT.
Two particularly tractable cases are studied in detail for which we can obtain renormalisation 
group flows for the parameters of the model together with an analysis of fixed point structure.
Of course these results are consistent with previous studies but we are not constrained to treat
Lorentz symmetry breaking as necessarily weak. As we found in a previous study of a scalar
field theory model an acceptable causal structure for the model imposes constraints on
relationship between the various lightcones in BIMQED.
}

\vfill
DAMTP-2016-26
\pagebreak

\section{\bf Introduction}

In a previous paper \cite{ITD2} we studied a scalar field theory model in which each field
was associated with a distinct metric. For simplicity we assumed that spacetime 
was represented by a flat background and parametrised by coordinates $x^\mu$
such that the theory is invariant under the translations $x^\mu\rightarrow x^\mu+a^\mu$.
The metrics are then independent of the coordinates $x^\mu$. We formulated the theory so 
that it was invariant under general linear transformations $x^\mu\rightarrow M^\mu_{~~\nu}x^\nu$.
The implications of the model were that in addition to a renormalisation of the 
coupling constants and masses such theories required a renormalisation of the metrics.
An examination of the renormalisation group showed that the important effect was
a renormalisation of the {\it relationships} between the  metrics and their associated lightcones.
These relationships are therefore dependent on the renormalisation scale $\mu$ (we 
use dimensional regularisation). We found that at each stage the associated lightcones must
overlap by sharing some interior vectors that are timelike in all the metrics. This constraint
on the lightcones originates in a requirement that the evolution of the full system
of interacting quantum fields is causal for some set of observers. The renormalisation of the metric relationship 
will have implications for models in which two (or more) metrics become dynamical degrees
of freedom.

Of course such models exhibit a breakdown of Lorentz invariance and correspond to a subset
of $CPT$-even violations. A more general set of effects has been the focus of a wide range 
of investigations by many authors \cite{KOST1,KOST2,KOST3,KOST4,COLGL1,COLGL2}. We believe however that 
our approach of concentrating on multimetric theories sheds further light on this particular sector of 
the violation of Lorentz invariance.

In this paper we consider a version of QED, Bimetric QED (BIMQED) that associates one metric with 
the electromagnetic and another with the electron field. A simple extension of the model could
involve the introduction of a third metric associated with the muon field, though we do not
pursue this here. Of course we do not see this model as any more than a demonstration of the ideas 
in a simple gauge theory since there is so far no observational reason for anticipating a breakdown 
of Lorentz invariance in QED. The same analysis applied to theories with non-abelian gauge groups 
is also of interest especially in relation to high energy scattering. In fact the model is related to, 
but in some ways simpler than, those investigated by Nielsen and Ninomiya \cite{NLSN1}, and  subsequently by Nielsen
and Picek \cite{NLSN2} and Chadha and Nielsen \cite{NLSN3}.

Our starting point is a formulation of electrodynamics that has been referred to as "pre-metric" \cite{ITIN}.
We examine this formulation and show that in fact there is a {\it preferred} metric. Identifying this metric
also permits a clear statement of the nature of Lorentz symmetry breaking for electrodynamics.
The breaking of Lorentz symmetry is associated with a tensor that has the symmetry properties
of the Weyl tensor in general relativity. We refer to this as a Weyl-like tensor (WLT). 
The Petrov classification \cite{PTRV} for such tensors can be used to identify the different kinds of symmetry 
breaking that are possible. Of course this analysis is consistent with other work \cite{KLINK,SCHRK,LEHN} on 
the breaking of Lorentz symmetry in electrodynamics. A feature of our approach is that while we do use the standard 
perturbation expansion in electric charge we are not constrained to treat Lorentz symmetry breaking
perturbatively, unless this happens to be convenient.

\section{\label{PREFM} Preferred Metric in Electrodynamics}

The pre-metric formulation of electrodynamics \cite{ITIN} in its most general form replaces
Maxwell's equations for the gauge field $A_\mu(x)$ with a modified set of the form
\be
\d_\mu \Ut^{\mu\nu\ss\tau}F_{\ss\tau}(x)=0,
\label{GINVEQM1}
\ee
where 
\be
F_{\ss\tau}(x)=\d_\ss A_\tau(x)-\d_\tau A_\ss(x),
\label{EMT}
\ee
and the (constant) tensor density $\Ut^{\mu\nu\ss\tau}$ satifies
\be
\Ut^{\mu\nu\ss\tau}=-\Ut^{\nu\mu\ss\tau}=-\Ut^{\mu\nu\tau\ss}.
\ee
In this general formulation there is no requirement that $\Ut^{\mu\nu\ss\tau}=\Ut^{\ss\tau\mu\nu}$.
Indeed the non-vanishing of the antisymmetric contribution $\Ut^{\mu\nu\ss\tau}-\Ut^{\ss\tau\mu\nu}$
produces what are referred to as skewon effectsi \cite{ITIN}. Since however we wish to derive our 
dynamical equations from a Lagrangian formulation we will exclude skewon effects. We will also
introduce a (constant) metric $g_{\mu\nu}$  and set
\be
\Ut^{\mu\nu\ss\tau}=\Om U^{\mu\nu\ss\tau},
\ee
where
\be
\det g_{\mu\nu}=-\Om^2,
\label{DET1}
\ee
and the tensor $U^{\mu\nu\ss\tau}$ satisfies
\begin{eqnarray}
U^{\mu\nu\ss\tau}&=&-U^{\nu\mu\ss\tau},\\\nonumber
U^{\mu\nu\ss\tau}&=&-U^{\mu\nu\tau\ss},\\\nonumber
U^{\mu\nu\ss\tau}&=&U^{\ss\tau\mu\nu},
\end{eqnarray}
together with
\be
U^{\mu\nu\ss\tau}+U^{\mu\ss\tau\nu}+U^{\mu\tau\nu\ss}=0.
\ee
As remarked by Nielsen and Ninomiya \cite{NLSN1}, this gives $U^{\mu\nu\ss\tau}$ the algebraic properties
of the Riemann tensor although there is no necessary connection. However see ref \cite{ITDSJH,SHHO1,SHHO2,SHHO3}.
For the given metric there is a standard decomposition of $U^{\mu\nu\ss\tau}$ in the form
\be
U^{\mu\nu\ss\tau}=\fr{1}{12}U(g^{\mu\ss}g^{\nu\tau}-g^{\mu\tau}g^{\nu\ss})
                           +\fr{1}{2}(g^{\mu\ss}S^{\nu\tau}+S^{\mu\ss}g^{\nu\tau}-g^{\mu\tau}S^{\nu\ss}-S^{\mu\tau}g^{\nu\ss})
                           -C^{\mu\nu\ss\tau},
\ee
where
\begin{eqnarray}
S^{\mu\ss}&=&U^{\mu\ss}-\fr{1}{4}Ug^{\mu\ss},\\\nonumber
U^{\mu\ss}&=&g_{\nu\tau}U^{\mu\nu\ss\tau},\\\nonumber
U&=&g_{\mu\ss}U^{\mu\ss},
\end{eqnarray}
and 
\be
g_{\nu\tau}C^{\mu\nu\ss\tau}=0.
\ee
Clearly $g_{\mu\ss}S^{\mu\ss}=0$ and $C^{\mu\nu\ss\tau}$ has the algebraic properties of the Weyl tensor.

So far the metric is arbitrary. We can arrive at a preferred metric by demanding that
the tensor $U^{\mu\nu\ss\tau}$ is most nearly like $g^{\mu\ss}g^{\nu\tau}-g^{\mu\tau}g^{\nu\ss}$.
We implement this idea by requiring that the overlap amplitude $U^{\mu\nu\ss\tau}(g_{\mu\ss}g_{\nu\tau}-g_{\mu\tau}g_{\nu\ss})$
is stationary with respect to variations of $g_{\mu\ss}$ subject to the constraint in eq(\ref{DET1}).
Introducing the Lagrange multiplier $\ll$ and setting 
\be
\Fc=U^{\mu\nu\ss\tau}(g_{\mu\ss}g_{\nu\tau}-g_{\mu\tau}g_{\nu\ss})-\ll\det g_{\mu\ss}
\ee
we require that
\be
\fr{\d \Fc}{g_{\mu\ss}}=0.
\ee
This yields
\be
U^{\mu\ss}+\ll\Om^2g^{\mu\ss}=0.
\label{STMT}
\ee
It follows that for this {\it stationary} value of the metric that
\be
S^{\mu\ss}=0.
\ee
Of course we are assuming that $U^{\mu\nu\ss\tau}$ is such that it yields a unique solution of eq(\ref{STMT}) 
with the right type of (lightcone generating) metric.

The action for the elctromagnetic field that gives rise to eq(\ref{GINVEQM1}) is $S_{(p)}$ where
\be
S_{(p)}=-\fr{1}{8}\int d^4x\Om U^{\mu\nu\ss\tau}F_{\mu\nu}(x)F_{\ss\tau}(x).
\label{EMACT}
\ee
For a given metric we are free to adjust the normalisation of the field $A_\mu(x)$ so that
the normalisation of $U^{\mu\nu\ss\tau}$ is such that $U=12$. We have then $S_{(p)}$ is given by
eq(\ref{EMACT}) where
\be
U^{\mu\nu\ss\tau}=(g^{\mu\ss}g^{\nu\tau}-g^{\mu\tau}g^{\nu\ss})-C^{\mu\nu\ss\tau}.
\label{PRFMD}
\ee
It follows that when the Weyl-like tensor $C^{\mu\nu\ss\tau}$ vanishes $S_\gg$ is invariant under the
Lorentz group that leaves $g_{\mu\nu}$ invariant and there is no birefringence in the evolution of the 
elctromagnetic field. However when $C^{\mu\nu\ss\tau}$ is non-vanishing we do encounter birefringence and
the breaking of Lorentz invariance. The possible ways in which Lorentz symmetry breaking occurs can
therefore be given a Petrov classification appropriate to the tensor $C^{\mu\nu\ss\tau}$.

\section{\label{EQM} Equations of Motion}

The gauge invariant equations of motion, eq(\ref{GINVEQM1}), now take the form
\be
(g^{\mu\ss}g^{\nu\tau}-g^{\mu\tau}g^{\nu\ss}-C^{\mu\nu\ss\tau})\d_\ss\d_\mu A_\nu(x)=0.
\label{GINVEQM2}
\ee
This is the standard form with Lorentz symmetry breaking for electrodynamics.
Our analysis makes clear that the there is no lack of generality in this form
and that we can always choose the preferred metric for which the Lorentz 
symmetry breaking is described by the traceless WLT $C^{\mu\nu\ss\tau}$.

If we seek a solution of the form
\be
A_\nu(x)=\veps_\nu e^{-iq.x},
\ee
then $\veps_\nu$ satisfies
\be
M^{\tau\nu}\veps_\nu=0,
\label{GINVEQM3}
\ee
where
\be
M^{\tau\nu}=q^2g^{\tau\nu}-q^\tau q^\nu-C^{\mu\nu\ss\tau}q_\mu  q_\ss,
\label{GINVEQM4}
\ee
and we set $q^\mu=g^{\mu\nu}q_\nu$.
Of course eq(\ref{GINVEQM3}) has the solution $\veps_\mu\propto q_\mu$ that for any vlaue of $q_\mu$.
It corresponds to a gauge degree of freedom. The physical solutions appear only when $q_\mu$ is constrained
to satisfy a dispersion relation that permits the rank of the matrix $M^{\tau\nu}$ to drop below the value 3
and its kernel to have a dimension greater than 1. See also ref \cite{ITIN}.

Following conventional lines of reasoning
we can explore plane wave solutions by imposing the gauge condition
\be
q^\nu\veps_\nu=0.
\label{GC1}
\ee
Eq(\ref{GINVEQM3}) becomes
\be
\left\{q^2g^{\tau\nu}-C^{\mu\nu\ss\tau}q_\mu  q_\ss\right\}\veps_\nu=0.
\label{GINVEQM5}
\ee
It is easy to see that eq(\ref{GINVEQM5}) implies 
\be
q^2q^\nu\veps_\nu=0.
\ee
Hence eq(\ref{GC1}) can be imposed in a consistent manner. The problem then reduces to 
finding the conditions on $q_\mu$ that allow eq(\ref{GINVEQM5}) to have nontrivial solutions.
We will return to the issue of gauge conditions later.

\subsection{\label{NP} Newman-Penrose Tetrad}
 
It is convenient to reformulate these equations of motion in terms of a Newman-Penrose
tetrad \cite{NEWPEN,JMS}. It comprises four null vectors, $l_\mu$, $n_\mu$, $m_\mu$ and $\mb_\mu$
where $l_\mu$ and $n_\mu$ are real and $m_\mu$ and $\mb_\mu$ are complex conjugates.
They satisfy the relations
\be
l^\mu l_\mu=n^\mu n_\mu=m^\mu m_\mu=\mb^\mu \mb_\mu=0,
\ee
and 
\be
l^\mu m_\mu=l^\mu \mb_\mu=n^\mu m_\mu=n^\mu \mb_\mu=0,
\ee
together with
\be
l^\mu n_\mu=-m^\mu \mb_\mu=1.
\ee
We have also
\be
g^{\mu\nu}=l^\mu n^\nu+n^\mu l^\nu-m^\mu\mb^\nu-\mb^\mu m^\nu.
\ee
Hence $\veps$ can be re-expressed in terms of its components in the NP tetrad basis,
\be
\veps^\mu=l^\mu(n.\veps)+n^\mu(l.\veps)-m^\mu(\mb.\veps)-\mb^\mu(m.\veps).
\ee
For later convenience we reformulate eq(\ref{GINVEQM5}) also in the tetrad basis.
Introduce $N^{\nu\tau}$ where
\be
N^{\nu\tau}=C^{\mu\nu\ss\tau}q_\mu q_\ss,
\ee
and the matrix entries $N_{ll}=N^{\nu\tau}l_\nu l_\tau$, $N_{ln}=N^{\nu\tau}l_\nu n_\tau$ {\it etc}.
We have then
\be
\left(
\begin{array}{cccc}
q^2-N_{ln}&-N_{ll}&N_{l\mb}&N_{lm}\\
-N_{nn}&q^2-N_{nl}&N_{n\mb}&N_{nm}\\
-N_{mn}&-N_{ml}&q^2+N_{m\mb}&N_{mm}\\
-N_{\mb n}&-N_{\mb l}&N_{\mb\mb}&q^2+N_{\mb m}
\end{array}
\right)
\left(
\begin{array}{c}
l.\veps\\ n.\veps\\ m.\veps\\ \mb.\veps
\end{array}
\right)=0.
\label{GINVEQM6}
\ee
For non-trivial solutions we require the vanishing of the determinant of the matrix in eq(\ref{GINVEQM6}).
In examples below we will see that this determinant has a factor of $(q^2)^2$ corresponding to
gauge modes. The remaining factor yields the lightcone structure of the physical modes.

\section{\label{PETROV} Petrov Classification of Lorentz Symmetry Breaking}

The Petrov classification of Weyl-like tensors (WLTs) \cite{PTRV} can be expressed in a number of ways.
A powerful way of understanding the structure of WLTs is the Newman-Penrose (NP) formalism together
with the Penrose spinor approach \cite{PENRIN}. 
A succinct account of the this formalism is provided by Stewart \cite{JMS}. A simple account may also
be found in \cite{PODON}. An important concept in
relation to a WLT is that of a principal null direction (PND). Such a PND is represented
by a null vector, $v_\mu$, that satisfies the constraint 
\be
v_{[\aa}C_{\mu]\nu\ss[\tau}v_{\bb]}v^\nu v^\ss=0.
\label{PR1}
\ee
In general there are four distinct directions that are solutions of this equation and
the WLT is Petrov class I. When the constraint has one double root and there are three 
distinct PNDs the WLT is Petrov class II. The case of two double roots is Petrov class D.
The PND corresponding to a double root in these two cases satisfies a modified 
(but consistent) constraint 
\be
C_{\mu\nu\ss[\tau}v_{\bb]}v^\nu v^{\ss}=0.
\label{PR2}
\ee
When there is a triple root, and two distinct PNDs, the WLT is of Petrov class III
and the PND corresponding to the triple root satisfies a further modified constraint
\be
C_{\mu\nu\ss[\tau}v_{\bb]}v^\ss=0.
\label{PR3}
\ee
Finally when all four roots coincide the WLT is of Petrov class N and the PND
satisfies the constraint
\be
C_{\mu\nu\ss\tau} v^\ss=0.
\label{PR4}
\ee
The underlying algebra that supports these results together with further implications 
utilises the NP and spinor formalism that is explained in refs \cite{NEWPEN,PENRIN,JMS,PODON}.

\subsection{\label{CANP} Canonical Forms for the Weyl-like Tensor}

It is useful for the purposes of explicit calculation to identify canonical forms 
associated with the Petrov classification of the WLTs. While these are not unique 
they may be expressed in terms of the NP tetrad.
It is convenient to introduce the antisymmetric tensors, $A_{\mu\nu}$, $B_{\mu\nu}$ and $D_{\mu\nu}$
where 
\begin{eqnarray}
A_{\mu\nu}&=&l_\mu m_\nu-l_\nu m_\mu\nonumber\\
B_{\mu\nu}&=&\mb_\mu n_\nu-\mb_\nu n_\mu\nonumber\\
D_{\mu\nu}&=&l_\mu n_\nu-l_\nu n_\mu+\mb_\mu m_\nu-\mb_\nu m_\mu
\end{eqnarray}
For class N we can choose the single PND to be $l_\mu$ and the WLT to
have the form
\be
C_{\mu\nu\ss\tau}=A_{\mu\nu}A_{\ss\tau}+\mbox{c.c.},
\ee
where $\mbox{c.c.}$ indicates complex conjugate. 
For class III we have
\be
C_{\mu\nu\ss\tau}=A_{\mu\nu}D_{\ss\tau}+D_{\mu\nu}A_{\ss\tau}+\mbox{c.c.}.
\ee
For class D 
\be
C_{\mu\nu\ss\tau}=\ll\{A_{\mu\nu}B_{\ss\tau}+B_{\mu\nu}A_{\ss\tau}
                                 +D_{\mu\nu}D_{\ss\tau}\} +\mbox{c.c.}.
\ee
For class II
\be
C_{\mu\nu\ss\tau}=\ll\{A_{\mu\nu}A_{\ss\tau}+\fr{1}{6}[A_{\mu\nu}B_{\ss\tau}+B_{\mu\nu}A_{\ss\tau}
                                +D_{\mu\nu}D_{\ss\tau}]\}+\mbox{c.c.}.
\ee
For class I
\be
C_{\mu\nu\ss\tau}=\mu\{A_{\mu\nu}A_{\ss\tau}+B_{\mu\nu}B_{\ss\tau}\}
       +\ll\{A_{\mu\nu}B_{\ss\tau}+B_{\mu\nu}A_{\ss\tau}+D_{\mu\nu}D_{\ss\tau}\}+\mbox{c.c.}.
\ee
In classes I, II, and D, $\ll$ and $\mu$ are complex parameters. In classes III and N,
any such complex parameter can be absorbed into the definition of $l$ and $m$
by subjecting them to an appropriate Lorentz transformation and rotation respectively.
However in the context of renormalisation analysis this may not always be convenient.
Where appropriate we will reinstate coefficients.

\subsection{\label{PETO} Example for Petrov Class O}

The very simplest case of Petrov class O, in which $C^{\mu\nu\ss\tau}$ vanishes is
not included in the above list. For a non-interacting elctromagnetic field it implies
no Lorentz symmetry breakdown. However in the case of QED, Lorentz symmetry may be broken 
because of differing lightcone structure for the photons and the electrons without
the introducion of a WLT into the dynamics. We will study such cases later in the context of 
renormalisation theory.

\subsection{\label{PETN} Example for Petrov Class N}

The simplest non-trivial case is Petrov class N. Eq(\ref{GINVEQM6}) becomes
\be
\left(
\begin{array}{cccc}
q^2&0&0&0\\
-(m.q)^2-(\mb.q)^2&q^2&(l.q)(m.q)&(l.q)(\mb.q)\\
-(l.q)(\mb.q)&0&q^2&(l.q)^2\\
-(l.q)(m.q)&0&(l.q)^2&q^2
\end{array}
\right)\left(
\begin{array}{c}
l.\veps\\ n.\veps\\ m.\veps\\ \mb.\veps
\end{array}
\right)=0.
\label{GINVEQM7}
\ee
It is easy to show that the determinant $\Delta$ of the matrix in eq(\ref{GINVEQM7})
is given by
\be
\Delta=(q^2)^2((q^2)^2-(l.q)^4).
\label{DELPN1}
\ee
We have then either 
\be
q^2=0~~\mbox{(twice)}, 
\ee
or
\be
q^2=\pm (l.q)^2.
\label{DELPN2}
\ee
When $q^2=0$, $(n.\veps)$ is arbitrary and there remains the solution for which $\veps\propto q$.
The general solution is then
\be
\veps_\tau=\aa q_\tau+\bb l_\tau,
\ee
where $\aa$ and $\bb$ are arbitrary parameters.

When $q^2\ne 0$ we find $l.\veps=0$. Then we have
\begin{eqnarray}
q^2m.\veps+(l.q)^2\mb.\veps&=&0,\nonumber\\
(l.q)^2m.\veps+q^2\mb.\veps&=&0.
\end{eqnarray}
Eq(\ref{DELPN2}) then allows non-trivial solutions $\veps^{(\pm)}$ which
satisfy
\be
(m\pm\mb).\veps^{(\pm)}=0.
\label{DELPN3}
\ee
In the present example then we see that Lorentz symmetry breakdown is revealed
by the birefringence associated with the two lightcones implicit in eq(\ref{DELPN2}). 
The corresponding polarisation vectors are determined through eq(\ref{DELPN3})
by the spatial axes $m$ and $\mb$.

One further point of significance is that by making a Lorentz transformation of
coordinates in the $l$-$n$ plane with a hyperbolic angle $\psi$, 
the PND vector becomes $e^\psi l_\mu$. Correspondingly $n_\mu\rightarrow e^{-\psi}n_\mu$.
The size of the components of $l_\mu$, as remarked above, can therefore be 
adjusted arbitrarily simply by making an appropriate choice of $\psi$.
In a sense then, the same Lorentz symmetry breaking situation 
can be viewed as either large or small depending which coordinate basis is appropriate
for describing the motion of the relevant observer.

\subsection{\label{PETIII} Example for Petrov Class III}

When the WLT is Petrov class III there are two PNDs, a triple root $l_\mu$ and a single root $n_\mu$.
They can be embedded in the NP tetrad as above and used to construct the the WLT thus
\be
C_{\mu\nu\ss\tau}=A_{\mu\nu}D_{\ss\tau}+D_{\mu\nu}A_{\ss\tau}+\mbox{c.c.}.
\ee
We find that eq(\ref{GINVEQM6}) becomes
\be
\Mc\left(
\begin{array}{c}
l.\veps\\n.\veps\\m.\veps\\\mb.\veps
\end{array}
\right)=0,
\label{GINVEQM8}
\ee
where the columns, $\Mc_i$ $(i=1,2,3,4)$, of the matrix $\Mc$ are given by
\be
\Mc_1=
\left(
\begin{array}{c}
q^2+l.qm.q+l.q\mb.q\\
-2n.pm.q-2n.q\mb.q\\
-l.qn.q-m.q\mb.q+(m.q)^2\\
-l.qn.q-m.q\mb.q+(\mb.q)^2
\end{array}
\right)
\ee
\be
\Mc_2=
\left(
\begin{array}{c}
0\\
q^2+l.qm.q+l.q\mb.q\\
(l.q)^2\\
(l.q)^2
\end{array}
\right)
\ee
\be
\Mc_3=
\left(
\begin{array}{c}
-(l.q)^2\\
m.q\mb.q+l.qn.q-(\mb.q)^2\\
q^2-l.qm.q-l.q\mb.q\\
2l.q\mb.q
\end{array}
\right)
\ee
\be
\Mc_4=
\left(
\begin{array}{c}
-(l.q)^2\\
l.qn.q+m.q\mb.q-(m.q)^2\\
2l.qm.q\\
q^2-l.qm.q-l.q\mb.q
\end{array}
\right)
\ee
After some calculation the determinant of $\Mc$ can be obtained in the form
\be
\det\Mc=(q^2)^2[(q^2-(l.q)^2)^2-(4l.qm.q-(l.q)^2)(4l.q\mb.q-(l.q)^2)].
\label{DELPIII1}
\ee
As expected on general grounds, see ref \cite{ITIN}, the second factor in eq(\ref{DELPIII1})
is a quartic expression the vanishing of which  
yields the dispersion relations for the two physical photon modes. There are two branches
\be
q^2-(l.q)^2=\pm (l.q)\sqrt{(4m.q-l.q)(4\mb.q-l.q)}
\ee
However in contrast to the previous example for Petrov class N, the quartic does not have quadratic factors.
Therefore the birefringence in this case {\it cannot} be described by two distinct conventional lightcones.
As is implicit in the derivation, the absolute strength of $C_{\mu\nu\ss\tau}$ can can be adjusted by
changing coordinates through Lorentz boosts in the $l_\mu$-$n_\mu$ plane and rotations in the $m$-$\mb$ plane. 
In this way Petrov class III has features in common with class N.

\subsection{\label{PETD} Example for Petrov Class D}

When the WLT is Petrov class D, there are again two PNDs, $l_\mu$ and $n_\mu$ both 
being double roots. They can be embedded in the NP tetrad and yield a WLT of the form
\be
C_{\mu\nu\ss\tau}=\ll\{A_{\mu\nu}B_{\ss\tau}+B_{\mu\nu}A_{\ss\tau}
                         +D_{\mu\nu}D_{\ss\tau}\}+\mbox{c.c.}.
\ee
We find that eq(\ref{GINVEQM6}) takes the form eq(\ref{GINVEQM8}) where
the columns of $\Mc$ 
are (we denote the complex conjugate of $\ll$ by $\llb$)
\be
\Mc_1=
\left(
\begin{array}{c}
q^2+(\ll+\llb)(l.qn.q+m.q\mb.q)\\
-(\ll+\llb)(n.q)^2\\
(2\ll-\llb)n.qm.q\\
(2\llb-\ll)n.q\mb.q
\end{array}
\right)
\ee
\be
\Mc_2=
\left(
\begin{array}{c}
-(\ll+\llb)(l.q)^2\\
q^2+(\ll+\llb)(l.qn.q+m.q\mb.q)\\
(2\llb-\ll)l.qm.q\\
(2\ll-\llb)l.q\mb.q
\end{array}
\right)
\ee
\be
\Mc_3=
\left(
\begin{array}{c}
-(2\ll-\llb)l.q\mb.q\\
-(2\llb-\ll)n.q\mb.q\\
q^2-(\ll+\llb)(l.qn.q+m.q\mb.q)\\
(\ll+\llb)(\mb.q)^2
\end{array}
\right)
\ee
\be
\Mc_4=
\left(
\begin{array}{c}
-(2\llb-\ll)l.qm.q\\
-(2\ll-\llb)n.qm.q\\
(\ll+\llb)(m.q)^2\\
q^2-(\ll+\llb)(l.qn.q+m.q\mb.q)
\end{array}
\right)
\ee
In the present case the coefficient $\ll$ cannot be absorbed by a redefinition of the  
vectors of the NP tetrad. The determinant of $\Mc$ can be obtained in the form
\be
\det\Mc=(1+\ll+\llb)(q^2)^2\{(q^2-(\ll+\llb)l.qn.q+\rho m.q\mb.q)^2
                   -9\ll\llb(1+\kk)(1+\kkb)(m.q)^2(\mb.q)^2\},
\label{DELPD1}
\ee
where the parameters $\kk$ and $\rho$ are given by
\be
\kk=\fr{1-(2\ll-\llb)}{1+\ll+\llb}.
\ee
and
\be
\rho=\ll+\llb-\fr{9\ll\llb}{1+\ll+\llb}.
\ee
The dispersion relation for the physical modes implied by the vanishing of $\det \Mc$ 
does factorise in this case and yields two distinct light cones each with a dispersion
relation that is quadratic in $q$, 
\be
q^2-(\ll+\llb)l.qn.q+\rho m.q\mb.q=\pm 3\sqrt{\ll\llb(1+\kk)(1+\kkb)}m.q\mb.q.
\ee

\subsection{\label{PETII} Example for Petrov Class II}

When the WLT is Petrov class II, there are three PNDs. In terms of the
basis of the NP tetrad the double root is $l_\mu$ and 
the two single roots are $l_\mu+n_\mu\mp i(m_\mu-\mb_\mu)$.
The  WLT has the form
\be
C_{\mu\nu\ss\tau}=\ll\{A_{\mu\nu}A_{\ss\tau}+A_{\mu\nu}A_{\ss\tau}
                    +\fr{1}{6}[A_{\mu\nu}B_{\ss\tau}+B_{\mu\nu}A_{\ss\tau}
                                       `+D_{\mu\nu}D_{\ss\tau}]\}+\mbox{c.c.}.
\ee
We find that eq(\ref{GINVEQM6}) takes the form eq(\ref{GINVEQM7}) where
the columns of $\Mc$ are
\be
\Mc_1=
\left(
\begin{array}{c}
q^2+\fr{1}{6}(\ll+\llb)(l.qn.q+m.q\mb.q)\\
-\ll(m.q)^2-\llb(\mb.q)^2-\fr{1}{6}(\ll+\llb)(n.q)^2\\
-\llb l.q\mb.q+\fr{1}{6}(2\ll-\llb)n.qm.q\\
-\ll l.qm.q+\fr{1}{6}(2\llb-\ll)n.q\mb.q
\end{array}
\right)
\ee
\be
\Mc_2=
\left(
\begin{array}{c}
-\fr{1}{6}(\ll+\llb)(l.q)^2\\
q^2+\fr{1}{6}(\ll+\llb)(l.qn.q+m.q\mb.q)\\
\fr{1}{6}(2\llb-\ll)l.qm.q\\
\fr{1}{6}(2\ll-\llb)l.q\mb.q
\end{array}
\right)
\ee
\be
\Mc_3=
\left(
\begin{array}{c}
-\fr{1}{6}(2\ll-\llb)l.q\mb.q\\
\ll l.qm.q-\fr{1}{6}(2\llb-\ll)n.q\mb.q\\
q^2-\fr{1}{6}(\ll+\llb)(l.qn.q+m.q\mb.q)\\
\ll(l.q)^2+\fr{1}{6}(\ll+\llb)(\mb.q)^2
\end{array}
\right)
\ee
\be
\Mc_4=
\left(
\begin{array}{c}
-\fr{1}{6}(2\llb-\ll)l.qm.q\\
\llb l.q\mb.q-\fr{1}{6}(2\ll-\llb)n.qm.q\\
\llb(l.q)^2+\fr{1}{6}(\ll+\llb)(m.q)^2\\
q^2-\fr{1}{6}(\ll+\llb)(l.qn.q+m.q\mb.q)
\end{array}
\right)
\ee
We find that
\begin{eqnarray}
\det\Mc&=
&[(1+\fr{1}{6}(\ll+\llb)](q^2)^2\{[(1-\fr{1}{12}(\ll+\llb))q^2
                    +\fr{1}{2}\llb(\kk-1)m.q\mb.q]~~~~~~~~~\\\nonumber
&&~~~~~~~~~~~~~~~~~~~~~~~~~~~~~\times[(1-\fr{1}{12}(\ll+\llb))q^2+\fr{1}{2}\ll(\kkb-1)m.q\mb.q]\\\nonumber
&&~~~~~~~~~~~~~~~~~-[\ll(l.q)^2+\fr{1}{2}\ll(\kk+1)(\mb.q)^2][\llb(l.q)^2+\fr{1}{2}\llb(\kkb+1)(m.q)^2]\}.
\label{DELPII1}
\end{eqnarray}
Here we have
\be
\kk=\fr{1-(2\ll-\llb)/6}{1+(\ll+\llb)/6}
\ee
Clearly the quartic expression providing the dispersion relation for the physical modes
again does not, in general, factorise into separate quadratic factors so the lightcone structure is
more complex than two separate simple lightcones. A special case is $\ll$ real when 
factorisation does occur.

\subsection{\label{PETI} Example for Petrov Class I}

When the WLT is Petrov class I, there are four PNDs. A canonical form
for the WLT in terms of the basis of the NP tetrad is
\be
C_{\mu\nu\ss\tau}=\mu\{A_{\mu\nu}A_{\ss\tau}+B_{\mu\nu}B_{\ss\tau}\}
                          + \ll\{A_{\mu\nu}B_{\ss\tau}+B_{\mu\nu}A_{\ss\tau}
                            +D_{\mu\nu}D_{\ss\tau}\}+\mbox{c.c.}.
\ee
We find that eq(\ref{GINVEQM6}) takes the form eq(\ref{GINVEQM7}) where
the columns of $\Mc$ are
\be
\Mc_1=
\left(
\begin{array}{c}
q^2+(\ll+\llb)(l.qn.q+m.q\mb.q)\\
-\mu(m.q)^2-\mub(\mb.q)^2-(\ll+\llb)(n.q)^2\\
-\mub l.q\mb.q+(2\ll-\llb)n.qm.q\\
-\mu l.qm.q+(2\llb-\ll)n.q\mb.q
\end{array}
\right)
\ee
\be
\Mc_2=
\left(
\begin{array}{c}
-\mu(\mb.q)^2-\mub(m.b)^2-(\ll+\llb)(l.q)^2\\
q^2+(\ll+\llb)(l.qn.q+m.q\mb.q)\\
-\mu n.q\mb.q+(2\llb-\ll)l.qm.q\\
-\mub n.qm.q+(2\ll-\llb)l.q\mb.q
\end{array}
\right)
\ee
\be
\Mc_3=
\left(
\begin{array}{c}
\mub n.qm.q-(2\ll-\llb)l.q\mb.q\\
\mu l.qm.q-(2\llb-\ll)n.q\mb.q\\
q^2-(\ll+\llb)(l.qn.q+m.q\mb.q)\\
\mu(l.q)^2+\mub(n.q)^2+(\ll+\llb)(\mb.q)^2
\end{array}
\right)
\ee
\be
\Mc_4=
\left(
\begin{array}{c}
\mu n.q\mb.q-(2\llb-\ll)l.qm.q\\
\mub l.q\mb.q-(2\ll-\llb)n.qm.q\\
\mu(n.q)^2+\mub(l.q)^2+(\ll+\llb)(m.q)^2\\
q^2-(\ll+\llb)(l.qn.q+m.q\mb.q)
\end{array}
\right)
\ee
The determinant of $\Mc$ is 
\be
\det \Mc=(q^2)^2\{\DD_0+\mu\DD_1+\mub\DD_2+\mu^2\DD_3+\mub^2\DD_4
                       +\mu\mub\DD_5+\mu^2\mub\DD_6+\mu\mub^2\DD_7\},
\label{DELPI1}
\ee
where
\begin{eqnarray}
\DD_0&=&\fr{1}{4}(\ll+\llb+1)(\ll+\llb-2)(q^2)^2+18\ll\llb l.qn.qm.q\mb.q,\nonumber\\
\DD_1&=&-3\llb(2\ll-\llb-2)[(l.q)^2(m.q)^2+(n.q)^2(\mb.q)^2],\nonumber\\
\DD_2&=&-3\ll(2\llb-\ll-2)[(l.q)^2(\mb.q)^2+(n.q)^2(m.q)^2],\nonumber\\
\DD_3&=&-[(\ll+\llb+1)((l.q)^2(n.q)^2+(m.q)^2(\mb.q)^2)+2(2\llb-\ll-1)l.qn.qm.q\mb.q],\nonumber\\
\DD_4&=&-[(\ll+\llb+1)((l.q)^2(n.q)^2+(m.q)^2(\mb.q)^2)+2(2\ll-\llb-1)l.qn.qm.q\mb.q],\nonumber\\
\DD_5&=&-(\ll+\llb+1)[(l.q)^4+(n.q)^4+(m.q)^4+(\mb.q)^4],\nonumber\\
\DD_6&=&-[(l.q)^2(\mb.q)^2+(n.q)^2(m.q)^2],\nonumber\\
\DD_7&=&-[(l.q)^2(m.q)^2+(n.q)^2(\mb.q)^2].
\end{eqnarray}
The quartic factor in eq(\ref{DELPI1}) yields the dispersion relation for the physical modes.
Unsurprisingly it does not exhibit any obvious factorisation properties. Hence we expect  
for this case also there are no simple lightcones controlling photon propagation. There may be 
special choices for the parameters that does allow factorisation.

\section{\label{GENGF}Generalised Gauge Fixing for the EM Field}

In the above discussion we obtained the dispersion relations for physical modes
by imposing the gauge condition $g^{\mu\nu}\d_\mu A_\nu(x)=0$. However
in anticipation of issues that arise in the context of the renormalisation of gauge theories 
with a multi-metric structure we examine a more general form of gauge fixing for the 
electromagnetic (EM) field. We introduce a metric-like object $\LL^{\mu\nu}$ and
impose the gauge condition
\be
\LL^{\mu\nu}\d_\mu A_\nu(x)=0.
\ee
In appendix \ref{GFIX2} we present the standard argument using the fuctional formalism
to derive the gauge fixed action for the EM field. It is
\be
S_{(p)}=\int d^nx \{-\fr{1}{8}U^{\mu\nu\ss\tau}F_{\mu\nu}(x)F_{\ss\tau(x)}
           -\fr{1}{2}\LL^{\mu\nu}\LL^{\ss\tau}\d_\mu A_\nu(x)\d_\ss A_\tau(x)
                        -\d_\mu\cb(x)\LL^{\mu\nu}\d_\nu c(x)\}.
\ee 
Here $c(x)$ and $\cb(x)$ are the ghost fields. 
Because we intend to use dimensional regularisation we express our results in $n$ dimensions.
The argument identifying the preferred metric generalises straightforwardly to $n$ dimensions,
hence we can express $U^{\mu\nu\ss\tau}$ using eq(\ref{PRFMD}) interpreted in $n$ dimensions.

The equation of motion for the photon field is then
\be
(g^{\mu\ss}g^{\nu\tau}-(g^{\mu\tau}g^{\nu\ss}-\LL^{\mu\nu}\LL^{\ss\tau})
                                         -C^{\mu\nu\ss\tau})\d_\ss\d_\mu A_\nu(x)=0,
\label{EQMOTPH}
\ee
and those for the ghost fields are
\begin{eqnarray}
\LL^{\mu\nu}\d_\mu\d_\nu c(x)&=&0,\nonumber\\
\LL^{\mu\nu}\d_\mu\d_\nu\cb(x)&=&0.
\label{EQMOTGST}
\end{eqnarray}
Clearly $\LL^{\mu\nu}$ plays the role of the (inverse) metric for the ghost fields.
It follows that the null mass-shell condition for the ghosts is
\be
\LL^{\mu\nu}q_\mu q_\nu=0.
\ee
Even in the case of no Lorentz symmetry breaking this null mass-shell is distinct 
from the photon null mass-shell unless we set $\LL^{\mu\nu}=g^{\mu\nu}$.
In QED, of course the ghosts do not interact with the photons. For this reason the ghosts 
are usually ignored in QED calculations. We will keep them in mind in particular because they 
play a more significant role in  the corresponding situation in non-abelian gauge theories.
The presence of multiple null mass-shells or multiple lightcones in the theory raises the same 
same issues dealt with in a previous paper discussing a bimetric model with scalar fields. For the 
moment we will restrict our observations to the requirement that the parameters of the theory
should be constrained so that there exist foliations of spacetime that are spacelike with respect to 
all relevant metrics in order to permit a causal structure in the theory including
the ghosts \cite{ITD2}. 

\subsection{\label{PHWFS} Photon Wavefunctions}

Plane wave solutions of eq(\ref{EQMOTPH}) have the form
\be
A_\nu(x)=\veps_\nu e^{-iq.x},
\ee
where $\veps_\nu$ satisfies  
\be
M^{\tau\nu}\veps_\nu=0,
\label{WF1}
\ee
with 
\be
M^{\tau\nu}=\Mc^{\tau\nu}-q^\tau q^\nu+Q^\tau Q^\nu,
\ee
where we have set $Q^\mu=\LL^{\mu\nu}q_\nu$ and where
\be
\Mc^{\tau\nu}=q^2g^{\tau\nu}-N^{\tau\nu}.
\ee
Recall $N^{\tau\nu}=N^{\nu\tau}=C^{\mu\nu\ss\tau}q_\mu q_\ss$.
In this notation the ghost mass shell condition is
\be
q.Q=0.
\ee
We have also
\be
N^{\mu\nu}q_\nu=q_\mu N^{\mu\nu}=0.
\ee
We then find easily that eq(\ref{WF1}) implies
\be
q.QQ.\veps=0.
\ee
Provided that $q$ does not lie on the ghost mass shell the photon wavefunction satisfies
\be
Q.\veps=0,
\ee
which is precisely the gauge condition we wish to impose.
In order that eq(\ref{WF1}) have a solution it is necessary that $M^{\tau\nu}$ be singular. The inverse
of $M^{\tau\nu}$ is $M_{\nu\ll}$ satisfying $M^{\tau\nu}M_{\nu\ll}=\dd^\tau_\ll$ and has the form
\be
M_{\tau\nu}=\left(\dd^\rho_\tau-\fr{q_\tau Q^\rho}{Q.q}\right)\Mc_{\rho\ll}\left(\dd^\ll_\nu-\fr{Q^\ll q_\nu}{Q.q}\right)
                            +\fr{q_\tau q_\nu}{(Q.q)^2},
\label{WF2}
\ee
where $\Mc_{\rho\ll}$ is the inverse of $\Mc^{\rho\ll}$.

Clearly we can expect $M^{\tau\nu}$ to be singular when either
\be
Q^\mu q_\mu=\LL^{\mu\nu}q_\mu q_\nu=0,
\ee
or $\Mc^{\rho\ll}$ is singular. 
That is either $q_\mu$ lies on the ghost mass-shell or $q_\mu$ satisfies $\det\Mc^{\rho\ll}=0$. 
There may be special cases when these two constraints intersect and $q_\mu$ lies
in the intersection. For simplicity of exposition we will not consider these special cases explicitly.
The former condition above signals the presence of ghost contributions to the photon propagator.
This is not different from the standard case except that our gauge condition separates the ghost mass-shell 
from that of the physical modes. The physical modes are associated with the vanishing of $\det\Mc^{\rho\ll}$.
However as we have seen from our examination of the various Petrov classes $\det\Mc^{\rho\ll}$, in four
dimensions, contains a factor $(q^2)^2$. This remains true in $n$ dimensions. Nevertheless $q^2=0$ does 
not correspond to a singularity of $M^{\nu\tau}$. We show this explicitly in appendix \ref{FALSING}.

The relationship between elements of the kernals of $M^{\tau\nu}$ and $\Mc^{\tau\nu}$ can be exhibited 
directly. Let $\veps'$ satisfy
\be 
\Mc^{\tau\nu}\veps'_\nu=0.
\ee
It follows, assuming $q^2\ne 0$, that $q.\veps'=0$. Now introduce $\veps$ where
\be
\veps=\veps'+\aa q,
\ee
$\aa$ being chosen so that $Q.\veps=0$. It is then easy to check that
\be
M^{\tau\nu}\veps_\nu=0.
\ee
This relationship, a gauge transformation of course, explains why the
same physical dispersion relation emerges from either method of fixing the gauge.
Because a factor of $(q^2)^2$ can be extracted from $\det \Mc^{\tau\nu}$ 
it follows that the vanishing of the remaining factor is homogeneous in $q$
of degree $2(n-2)$ \cite{ITIN}. Confining attention to positive energy solutions we can 
expect in general $n-2$ branches for the physical disperion relation. As we have seen 
from our analysis of the Petrov classification, these may or may not factorise into separate
lightlike mass-shells. 

There is also a solution for which $\veps\propto q$, when $q$ lies on the ghost mass-shell.
The final contribution to the full suite of $n$ solutions is not pure plane wave but contains a secular term. 
It is not unique but can be chosen to have the form 
\be
A_\tau(x)=(a_\tau+iq_\tau x^0)e^{-iq_\mu x^\mu},
\ee
Here $q_\tau$ lies on the ghost mass-shell. The requirement that the above wavefunction
is a solution of eq(\ref{EQMOTPH}) only fixes $a_\tau$ up to a gauge transformation $a_\tau\rightarrow a_\tau+\aa q_\tau$.
It is convenient to complete the determination of $a_\tau$ by choosing $\aa$ so that $q.a=0$. 
An analogous solution appears in conventional gauge fixing in standard gauge theories
where it can be understood as the limit of a difference of two coinciding solutions in
the Stuckelberg approach to gauge theories.

\subsection{\label{PHGF} Photon Green's Functions}

We can compute the photon Green's functions from the generating functional, $Z_{(p)}[J]$,
where
\be
Z_{(p)}[J]=\int d[A]\exp\left\{iS_{(p)}+i\int d^nx A_\mu(x)J^\mu(x)\right\},
\ee
and
\be
S_{(p)}=\int d^nx \{-\fr{1}{8}U^{\mu\nu\ss\tau}F_{\mu\nu}(x)F_{\ss\tau}(x)
           -\fr{1}{2}\LL^{\mu\nu}\LL^{\ss\tau}\d_\mu A_\nu(x)\d_\ss A_\tau(x)\}.
\label{PHACTION}
\ee
Here $S_{(p)}$ is the gauge fixed action without the ghosts which do not play a role in QED calculations.

On completing the square in the exponent in the functional integral we find that
\be
Z_{(p)}[J]=C\exp\left\{i\int d^nxd^nx'(-\fr{1}{2}J^\ss(x)N_{\ss\tau}(x-x')J^\tau(x'))\right\},
\label{GENF4}
\ee
where
\be
N_{\tau\ll}(x-x')=-\int\fr{d^nq}{(2\pi)^n}e^{-iq_\mu(x-x')^\mu}M_{\tau\ll}(q),
\ee
The external factor $C=Z_{(p)}[0]$ and is irrelevant to subsequent calculations. It will be omitted
from now on.
The two-point Green's function for free photons is $iG_{F\mu\nu}(x-x')$ where
\be
iG_{F\mu\nu}(x-x')=\fr{1}{Z[J]}\fr{\dd}{i\dd J^\mu(x)}\fr{\dd}{i\dd J^\nu(x)}Z[J]|_{J=0}
                  =\int \fr{d^nq}{(2\pi)^n} e^{-iq_\mu(x-x')^\mu}(-iM_{\mu\nu}(q)).
\ee

\subsection{\label{PHGFREP} Representation for the Photon Green's Function}

It is convenient, for later use in computing divergences, to construct a representation for the photon
Green's function. The essential step is to obtain a representation for the core quantity $\Mc_{\rho\ll}(q)$.
First we introduce the matrix $\Mc^\mu_{~~\ll}(q)$ given by
\be
\Mc^\mu_{~~\ll}=\Mc^{\mu\nu}(q)g_{\nu\ll}=q^2\dd^\mu_\ll-N^\mu_{~~\ll}(q),
\ee
where
\be
N^\mu_{~~\ll}(q)=N^{\mu\nu}(q)g_{\nu\ll}=C^{\mu\ss~~\tau}_{~~~~\ll}q_\ss q_\tau.
\ee
Formally we write
\be
\Mc^\mu_{~~\ll}(q)=(\Mcb(q))^\mu_{~~\ll},
\ee
and
\be
N^\mu_{~~\ll}(q)=(\Nbf(q))^\mu_{~~\ll},
\ee
hence, in matrix notation,
\be
\Mcb(q)=q^21-\Nbf(q).
\ee
We have then
\be
\Mc_{\nu\rho}(q)=g_{\nu\ll}(\Mcb^{-1}(q))^\ll_{~~\rho}.
\ee
We now introduce the identity
\be
\Mcb^{-1}(q)=-i\int_0^\infty du\exp\{iu(\Mcb(q)+i\veps)\}=-i\int_0^\infty du e^{iu(q^2+i\veps)}\exp\{-iu\Nbf(q)\}.
\ee
We set
\be
\exp\{-iu\Nbf(q)\}=\exp\{-i\Nbf(-i\d_z)\}e^{{i\sqrt u}q.z},
\ee
with the proviso that afterwards we set $z=0$.
We then have the result
\be
\Mcb^{-1}(q)=-i\exp\{-i\Nbf(-i\d_z)\}\int_0^\infty du e^{iu(q^2+q.z/\sqrt{u}+i\veps)}.
\label{INVREP}
\ee

We will find this result useful later. It has the advantage of exhibiting the 
dependence of the photon propagator on the WLT, $C^{\mu\nu\ss\tau}$.

\subsection{\label{EGF} Electron Green's Functions}

Because we wish to associate a new lightcone with the electron field we introduce
a vierbein $\eb^a_{~~\mu}$ which allows us to create a new metric $\gb_{\mu\nu}=\eta_{ab}\eb^a_{~~\mu}\eb^a_{~~\nu}$
and appropriate Dirac matrices $\ggb^\mu=\eb_{~~a}^\mu\gg^a$. Here $\gg^a$ are standard Dirac matrices
that satisfy $\{\gg^a,\gg^b\}=2\eta^{ab}$, hence
\be
\{\ggb^\mu,\ggb^\nu\}=2\gb^{\mu\nu}.
\ee
The volume element is again $d^nx$. This can be achieved by adjusting appropriately the normalisation of the 
electron field.
 
The action for the free electron field is $S_{(e)}$ where
\be
S_{(e)}=\int d^nx\psib(x)(i\ggb^\mu\d_\mu-m)\psi(x),
\label{EACTION}
\ee
and $m$ is the mass parameter of the electron.
The Green's functions for the elctron field are obtained from the generating functional $Z_{(e)}[\eta,\etab]$
where
\be
Z_{(e)}[\eta,\etab]=\int d[\psi]d[\psib]\exp\left\{iS_{(e)}+i\int d^nx(\etab(x)\psi(x)+\psib(x)\eta(x))\right\},
\ee
and $\eta(x)$ and $\etab(x)$ are anticommuting fields. Completing the square in the exponent in the functional integral
we find
\be
Z_{(e)}[\eta,\etab]=C' \exp\left\{-i\int d^nxd^nx'\etab(x)\DD(x-x')\eta(x')\right\},
\ee
where
\be
\DD(x-x')=\int \fr{d^nq}{(2\pi)^n}\fr{(\ggb^\mu q_\mu+m)e^{-iq_\mu(x-x')^\mu}}{\gb^{\mu\nu} q_\mu q_\nu-m^2+i\eps}.
\ee
The coefficient $C'$ can be omitted. We have
\be
iS_F(x-x')=\fr{1}{Z[J,\eta,\etab]}\fr{1}{i}\fr{\dd}{\dd \eta(x')}\fr{1}{i}\fr{\dd}{\dd \etab(x)}Z[J,\eta,\etab]|_{J=\eta=\etab=0},
\ee
giving the result
\be
iS_F(x-x')=i\int\fr{d^nq}{(2\pi)^n}\fr{(\ggb^\mu q_\mu+m)e^{-iq_\mu(x-x')^\mu}}{\gb^{\mu\nu}q_\mu q_\nu-m^2+i\eps}.
\ee

\section{\label{BASMOD} Bimetric QED}

We now arrange for the electrons to interact with the electromagnetic field. 
Of course we must allow for the renormalisation of the bare parameters of the theory.
We denote these bare parameters with a zero suffix. The bare electric charge is $-e_0$ and
the bare mass parameter is $m_0$. From our previous paper we can anticipate that the 
metrics in the theory will also require renormalisation so we make the replacements
$g_{\mu\nu}\rightarrow g_{0\mu\nu}$, $\eb^a_{~~\mu}\rightarrow \eb^a_{0~\mu}$ and $\ggb^\mu\rightarrow \ggb_0^\mu=\eb^\mu_{0~a}\gg^a$.
We also have $\gb_{\mu\nu}\rightarrow\gb_{0\mu\nu}=\eta_{ab}\eb^a_{0~\mu}\eb^b_{0~\nu}$. We must also
allow for the necessity of renormalising the the WLT representing Lorentz symmetry breaking for the photon field
and make the replacement $C^{\mu\nu\ss\tau}\rightarrow C^{\mu\nu\ss\tau}_0$ and hence the obvious
corresponding replacement $U^{\mu\nu\ss\tau}\rightarrow U^{\mu\nu\ss\tau}_0$.
In addition the gauge fixing metric is modified with $\LL^{\mu\nu}\rightarrow \LL_0^{\mu\nu}$, in order to 
accommodate possible renormalisation effects. 

We deviate a little from the approach of \cite{ITD2} by identifying the volume elements of the two metrics 
as $d^nx$ which then remains unrenormalized. Instead we 
we rely in a conventional way, on field renormalisations to render the Green's functions of the theory finite.
In \cite{ITD2} we absorbed this field renormalisation into the renormalisation of the volume elements. In fact a 
study of the renormalisation group in \cite{ITD2} showed that it may be factored out again. In the present model it
turns out that we cannot in any case, carry out this manoeuvre because of the photon action depends quadratically on the 
(inverse) metric.  We will therefore adopt the more conventional scheme and treat the field renormalisations 
independently of the metric.

The action for the electromagnetic field $A_\mu(x)$ is the gauge fixed action $S_{(p)}$ is now 
given by eq(\ref{PHACTION}) with the replacement of parameters by their bare versions.
We omit the ghost fields from now on since they play no role in QED calculations.  
The action for the electron field is $S_{(e)}$ is obtained similarly from eq(\ref{EACTION})
together with the inclusion of the elctromagnetic interaction to yield
\be
S_{(e)}=\int d^nx\psib(x)(i\ggb_0^\mu\d_\mu+e_0\ggb_0^\mu A_\mu(x)-m_0)\psi(x).
\ee
The total action for the theory is
\be
S=S_{(p)}+S_{(e)}.
\ee
The generating functional for the Green's functions is
\be
Z[J,\eta,\etab]=\int d[A]d[\psi]d[\psib]\exp\{iS+i\int d^nx(J^\mu(x)A_\mu(x)+(\etab(x)\psi(x)+\psib(x)\eta(x)))\}.
\label{GENF1}
\ee
When $e_0=0$ $S$ reduces to $S_f$ the free action and $Z[J,\eta,\etab]$ becomes the free-particle generating functional, 
$Z_f[J,\eta,\etab]$, where
\be
Z_f[J,\eta,\etab]=\int d[A]d[\psi]d[\psib]\exp\{iS_f+i\int d^nx(J^\mu(x)A_\mu(x)+(\etab(x)\psi(x)+\psib(x)\eta(x)))\}.
\label{GENF2}
\ee
The full generating functional, $Z[J,\eta,\etab]$, can be reconstructed from $Z_f[J,\eta,\etab]$ by
means of the standard formula
\be
Z[J,\eta,\etab]=\exp\left\{ie_0\int d^nx
         \fr{1}{i}\fr{\dd}{\dd \eta(x)}\ggb^\mu_0\fr{1}{i}\fr{\dd}{\dd J^\mu(x)}\fr{1}{i}\fr{\dd}{\dd\etab(x)}\right\}
                                        Z_f[J,\eta,\etab].
\label{GENF3}
\ee

\subsection{\label{GF}Green's Functions with Interaction}

The Green's functions that are important for a study of the renormalisation of the theory are
the two point Green's function for photons, $G_{\mu\nu}(x-x')$, where
\be
iG_{\mu\nu}(x-x')=\fr{1}{Z[J,\eta,\etab]}\fr{1}{i}\fr{\dd}{\dd J^\mu(x)}\fr{1}{i}\fr{\dd}{\dd J^\nu(x')}
                                          Z[J,\eta,\etab]|_{J=\eta=\etab=0}.
\label{GF1}
\ee
the two point Green's function for electrons, $S(x-x')$, where
\be
iS(x-x')=\fr{1}{Z[J,\eta,\etab]}\fr{1}{i}\fr{\dd}{\dd \eta(x')}\fr{1}{i}\fr{\dd}{\dd \etab(x)}Z[J,\eta,\etab]|_{J=\eta=\etab=0}.
\label{GF2}
\ee
and the vertex function $V_\mu(x,x',y)$, where
\be
iV_\mu(x,x',y)=\fr{1}{Z[J,\eta,\etab]}\fr{1}{i}\fr{\dd}{J^\mu(y)}\fr{1}{i}\fr{\dd}{\dd \eta(x')}\fr{1}{i}\fr{\dd}{\dd \etab(x)}
                                                                       Z[J,\eta,\etab]|_{J=\eta=\etab=0}.
\label{GF3}
\ee
The lowest approximation to $G_{\mu\nu}(x-x')$ is obtained by substituting $Z_f$ for $Z$ in eq(\ref{GF1}). We have
\be
iG_{F\mu\nu}(x-x')=\fr{1}{i}\int \fr{d^nq}{(2\pi)^n}
                                                e^{-iq_\mu(x-x')^\mu}M_{0\mu\nu}(q).
\label{GF4}
\ee
We have made the substitution $M_{\mu\nu}\rightarrow M_{0\mu\nu}$ to indicate that bare parameters are involved in the 
construction of $M_{0\mu\nu}$. 

In the same way we obtain the lowest approximation to $S(x-x')$. It is
\be
iS_F(x-x')=-\fr{1}{i}\int\fr{d^nq}{(2\pi)^n}\fr{(\ggb_0^\mu q_\mu+m_0)e^{-iq_\mu(x-x')^\mu}}{\gb_0^{\mu\nu}q_\mu q_\nu-m_0^2+i\eps}.
\label{GF5}
\ee
The lowest approximation to $iV_\mu(x,x',y)$ is
\be
iV_{F\mu}(x,x',y)= \int d^ny'iG_{F\mu\nu}(y-y')iS_F(x-y')(ie_0\ggb_0^\nu) iS_F(y'-x').
\label{GF6}
\ee

\subsection{\label{FEYN} Feynman Rules for Bimetric QED}

The Feynman rules for computing the perturbative expansions of Green's functions (in terms of bare parameters) can
be read off from the expansion for $Z[J,\eta,\etab]$ in eq(\ref{GENF3}) and the Green's function formulae in eqs(\ref{GF1}), (\ref{GF2}).
The Feynman diagrams are the conventional ones with the lines corresponding to $iG_{F\mu\nu}(x-x')$
for photons, $iS_F(x-x')$ for electrons and $ie_0\ggb_{0\mu}$ for each vertex. In momentum space a photon line with
momentum $q_\mu$ is associated with a factor
$$
iG_{F\mu\nu}(q)=-iM_{0\mu\nu}(q),
$$
and each electron line with
$$
iS_F(q)=i\fr{(\ggb_0^\mu q_\mu+m_0)}{\gb_0^{\mu\nu}q_\mu q_\nu-m_0^2+i\eps}.
$$
The vertex is $ie_0\ggb_{0\mu}$. Of course there is momentum conservation at each vertex and each loop momentum $q_\mu$ is
integrated with a measure
$$
\int\fr{d^nq}{(2\pi)^n}.
$$
Each electron loop has associated with it a factor of $(-1)$.

\subsection{\label{RENORM} Renormalisation}

We use dimensional regularisation with minimal subtraction to renormalise the theory. We introduce a scale $\mu$ so that the electron 
charge can be expressed in the form
\be
e_0=(\mu)^{(4-n)/2}e(1+\sum_{k=1}^\infty a^{(k)}(n)(e^2)^k).
\label{REN1}
\ee
The term in the sum that is $O(e^2)$ is simply a single pole at $n=4$. The renormalized charge $e$ is dimensionless.
Similarly the bare mass is expressed in the form
\be
m_0=m(1+\sum_{k=1}^\infty b^{(k)}(n)(e^2)^k),
\label{REN2}
\ee
where $m$ is the renormalized mass parameter. Again the term that is $O(e^2)$ is a simple pole at $n=4$.

The new aspect of the renormalisation procedure required for bimetric QED is that we must also 
allow for a renormalisation of the metrics thus
\be
g_0^{\mu\nu}=g^{\mu\nu}+\sum_{k=1}^\infty g^{(k)\mu\nu}(e^2)^k.
\label{REN3}
\ee
Similarly for the vierbein associated with the electron
\be
\eb_{0~a}^{\mu}=\eb_{~~a}^{\mu}+\sum_{k=1}^\infty \eb_{~~~~~a}^{(k)\mu}(e^2)^k.
\label{REN4}
\ee
The WLT term also requires renormalisation 
\be
C_0^{\mu\nu\ss\tau}=C^{\mu\nu\ss\tau}+\sum_{k=1}^\infty C^{(k)\mu\nu\ss\tau}(e^2)^k.
\label{REN5}
\ee
Finally
\be
\LL_0^{\mu\nu}=\LL^{\mu\nu}+\sum_{k=1}^\infty \LL^{(k)\mu\nu}(e^2)^k.
\label{REN6}
\ee
Again the terms of $O(e^2)$ contain a simple pole at $n=4$. Higher order terms contain poles of increasingly higher order.
Note that since we require $\det g_0^{\mu\nu}=\det g^{\mu\nu}=-1$, it follows that
\be
g_{\mu\nu}g^{(1)\mu\nu}=0.
\label{REN7}
\ee

The renormalisation procedure is carried out by inserting these parameter expansions into the bare perturbation
series and re-expanding in the renormalised charge $e$. The $n$-dependent coefficients of powers of $e^2$ are adjusted so that
there are sufficient cancellations of poles at $n=4$ that residual singular structure can be removed by appropriate
field renormalisation factors yielding finally Green's functions that are without divergent terms at $n=4$.

\section{\label{VACP} Vacuum Polarisation}

\begin{figure}[t]
   \centering
  \includegraphics[width=0.6\linewidth]{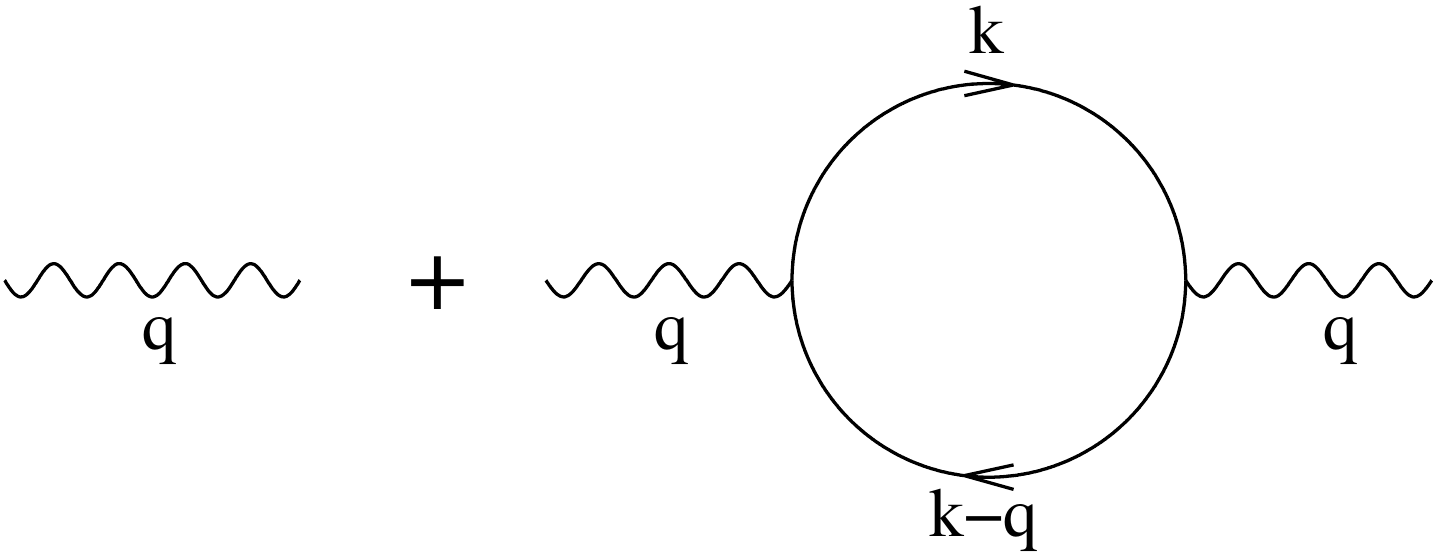}
  \caption{Contributions to the photon propagator to $O(e_0^2)$. }
  \label{FIG1}
\end{figure}

The lowest order contribution to the photon two-point function is represented by the Feynman 
diagram in Fig \ref{FIG1}.

Making use of the Feynman rules above we obtain to order $e_0^2$ 
\be
iG_{\mu\nu}(q)=iG_{F\mu\nu}(q)+iG_{F\mu\ss}(q)[i\SS^{\ss\tau}(q)]iG_{F\tau\nu}(q),
\label{VACP1}
\ee
where
\be
i\SS^{\ss\tau}(q)=(-1)(ie_0)^2\int \fr{d^nk}{(2\pi)^n}\Tr\{\ggb_0^\ss iS_F(k-q)\ggb_0^\tau iS_F(k)\ggb_0^\tau\}.
\label{VACP2}
\ee
If we introduce the inverse of the Green's function, $G^{\mu\nu}(q)$ where $G^{\mu\nu}(q)G_{\nu\ll}(q)=\dd^\mu_\ll$
and correspondingly $G_F^{\mu\nu}(q)$ where $G_F^{\mu\nu}(q)G_{F\nu\ll}(q)=\dd^\mu_\ll$ we find to second order
\be
G^{\mu\nu}(q)=G_F^{\mu\nu}(q)+\SS^{\mu\nu}(q).
\label{VACP3}
\ee
Since our calculation is an expansion to one loop we restrict the coupling constant expansion to $O(e^2)$. We have then  
\be
\SS^{\mu\nu}(q)=ie^2\mu^{(4-n)}\int\fr{d^nk}{(2\pi)^n}\fr{\Tr[\ggb^\mu(\ggb^\ll(k-q)_\ll+m)\ggb^\nu(\ggb^\rho k_\rho+m)]}
                                    {(\gb^{\aa\bb}(k-q)_\aa(k-q)_\bb-m^2+i\eps)(\gb^{\mu\nu}k_\mu k_\nu-m^2+i\eps)}.
\label{VACP4}
\ee
The calculation can be performed along essentially conventional lines. The divergent behaviour is
exhibited by calculating to $O(q^2)$ and we find
\be
\SS^{\mu\nu}(q)\simeq -\fr{e^2}{3}\fr{d(n)}{(4\pi)^{n/2}}\left(\fr{\mu}{m}\right)^{4-n} 
                            \GG(2-n/2)\left(\gb^{\mu\nu}\gb^{\aa\bb}-\gb^{\mu\bb}\gb^{\aa\nu}\right)q_\aa q_\bb.
\label{VACP5}
\ee
Here $d(n)$ is the dimension of the $\gg$-matrix representation. Of course $d(4)=4$. Therefore
we have a pole at $n=4$ of the form
\be
\SS^{\mu\nu}(q)\simeq \fr{e^2}{6\pi^2}\fr{1}{n-4}
                            \left(\gb^{\mu\nu}\gb^{\aa\bb}-\gb^{\mu\bb}\gb^{\aa\nu}\right)q_\aa q_\bb.
\label{VACP6}
\ee
Let the tensor $W^{\mu\aa\nu\bb}$ be given by
\be
W^{\mu\aa\nu\bb}=\gb^{\mu\nu}\gb^{\aa\bb}-\gb^{\mu\bb}\gb^{\aa\nu}.
\label{VACP7}
\ee
Since it has the same symmetry properties as $U^{\mu\aa\nu\bb}$, $W^{\mu\aa\nu\bb}$ can (in four dimensions)
be expressed in the form
\be
W^{\mu\aa\nu\bb}=\fr{1}{12}W(g^{\mu\nu}g^{\aa\bb}-g^{\mu\bb}g^{\aa\nu}) 
                     +\fr{1}{2}(V^{\mu\nu}g^{\aa\bb}+g^{\mu\nu}V^{\aa\bb}-V^{\mu\bb}g^{\aa\nu}-g^{\mu\bb}V^{\aa\nu})
                            -V^{\mu\nu\ss\tau},
\label{VACP8}
\ee
where
\begin{eqnarray}
W^{\mu\nu}&=&W^{\mu\aa\nu\bb}g_{\aa\bb},\nonumber\\
W&=&W^{\mu\nu}g_{\mu\nu},\nonumber\\
V^{\mu\nu}&=&W^{\mu\nu}-\fr{1}{4}Wg^{\mu\nu},\nonumber\\
V^{\mu\aa\ss\bb}g_{\aa\bb}&=&0.
\label{VACP9}
\end{eqnarray}
Here $V^{\mu\aa\nu\bb}$ is a WLT constructed ultimately from $\gb^{\mu\nu}$. Later we will examine
how different choices for $\gb^{\mu\nu}$ influence the Petrov class of $V^{\mu\aa\nu\bb}$. Of course
there is an $n$-dimensional version of this argument. However since we are simply calculating pole resiues 
at $n=4$ we will find here and later that the 4-dimensional calculation is sufficient.

From eq(\ref{GF4}) and eq(\ref{VACP3}) we see that
\be
G_F^{\mu\nu}(q)=-M_0^{\mu\nu}(q)+\SS^{\mu\nu}(q)
\ee
that is
\be
G_F^{\mu\nu}(q)=-(g_0^{\mu\nu}g_0^{\aa\bb}-g_0^{\mu\bb}g_0^{\aa\nu}+\LL_0^{\mu\bb}\LL_0^{\aa\nu}-C_0^{\mu\aa\nu\bb}
                                    -\fr{e^2}{6\pi^2}\fr{1}{n-4}W^{\mu\aa\nu\bb})q_\aa q_\bb
\ee
Using the expansions to $O(e^2)$ in eq(\ref{REN3}) to eq(\ref{REN5}) we see that we can remove
some of the UV poles at $n=4$ by choosing
\be
e^2g^{(1)\mu\nu}=\fr{e^2}{12\pi^2}\fr{1}{n-4}V^{\mu\nu},
\label{VACP10}
\ee
and
\be
e^2C^{(1)\mu\aa\nu\bb}=\fr{e^2}{6\pi^2}\fr{1}{n-4}V^{\mu\aa\nu\bb}-\fr{e^2}{72\pi^2}\fr{W}{n-4}C^{\mu\aa\nu\bb},
\label{VACP11}
\ee
with the result
\be
G_F^{\mu\nu}(q)=-\left\{\left(1-\fr{e^2}{72\pi^2}\fr{W}{n-4}\right)(g^{\mu\nu}g^{\aa\bb}-g^{\mu\bb}g^{\aa\nu}-C^{\mu\aa\nu\bb})
                                                      +\LL_0^{\mu\bb}\LL_0^{\aa\nu}\right\}q_\aa q_\bb.
\ee
This may be expressed in the form
\be
G_F^{\mu\nu}(q)=-\left(1-\fr{e^2}{72\pi^2}\fr{W}{n-4}\right)\left\{(g^{\mu\nu}g^{\aa\bb}-g^{\mu\bb}g^{\aa\nu})
                                                      +\LL^{\mu\bb}\LL^{\aa\nu}\right\}q_\aa q_\bb,
\ee
provided we arrange the expansion for the ghost mass-shell metric to satisfy
\be
\LL_0^{\mu\bb}=\left(1-\fr{e^2}{144\pi^2}\fr{W}{n-4}\right)\LL^{\mu\bb}.
\ee
The renormalised Green's function, $G_{RF}^{\mu\nu}$ is then obtained by means of the appropriate multiplicative photon 
wavefunction renormalisation yielding
\be
G_{RF}^{\mu\nu}(q)=-\left\{(g^{\mu\nu}g^{\aa\bb}-g^{\mu\bb}g^{\aa\nu})
                                                      +\LL^{\mu\bb}\LL^{\aa\nu}-C^{\mu\aa\nu\bb}\right\}q_\aa q_\bb.
\ee
The reason then that we introduced a distinct ghost mass-shell metric was to permit
this multiplicative renormalisation for the photon Green's function.

\section{\label{ELPROP} Electron Propagator}

\begin{figure}[t]
   \centering
  \includegraphics[width=0.6\linewidth]{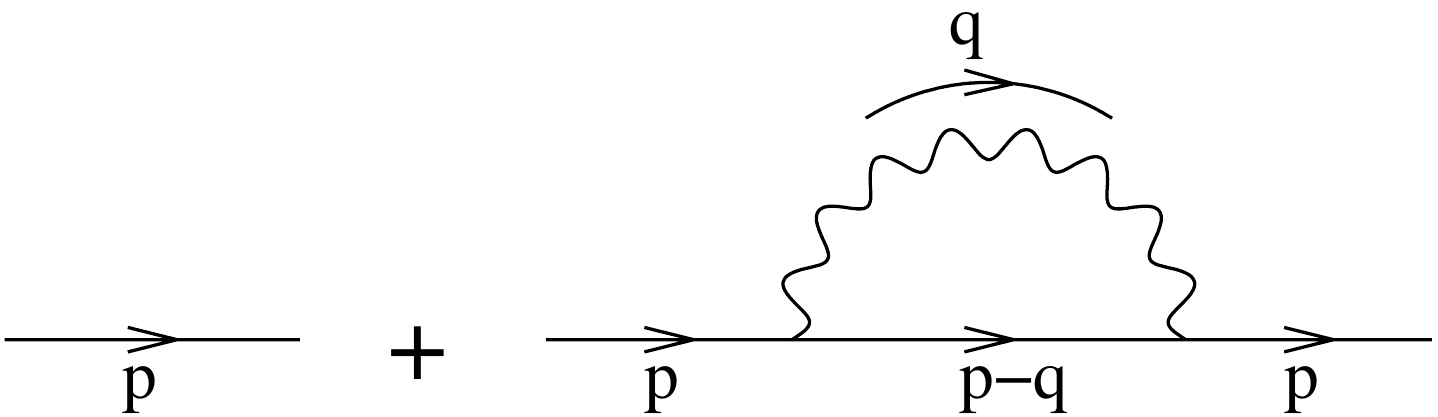}
  \caption{Contributions to the electron propagator to $O(e_0^2)$. }
  \label{FIG2}
\end{figure}

The lowest contributions to the electron propagator are shown in Fig \ref{FIG2}.

To $O(e_0^2)$ we have
\be
iS(p)=iS_F(p)+iS_F(p)i\SS(p)iS_F(p),
\ee
 where 
\be
i\SS(p)=(ie_0)^2\int\fr{d^nq}{(2\pi)^n}\ggb_0^\mu iS_F(p-q)\ggb_0^\nu(-iM_{0\mu\nu}(q)).
\ee
To second order we have
\be
S^{-1}(p)=S_F^{-1}(p)+\SS(p).
\label{PROPREN1}
\ee
Restricting the calculation to $O(e^2)$ we find
\be
\SS(p)=(ie)^2\mu^{4-n}\int\fr{d^nq}{(2\pi)^n}\ggb^\mu
             \fr{(\ggb^\ss(p_\ss-q_\ss)+m)}{\gb^{\aa\bb}(p_\aa-q_a)(p_\bb-q_\bb)-m^2+i\veps}\ggb^\nu(-iM_{\mu\nu}(q)).
\ee
The UV divergences in $\SS(p)$ are contained in the first two twrms of the Taylor series
\be
\SS(p)=\SS(0)+p_\mu\SS^\mu(0)+O(p^2),
\ee
where
\be
\SS^\mu(0)=\fr{\d}{\d p_\mu}\SS(p)|_{p=0}.
\ee
We have then
\be
\SS(0)=(ie)^2\mu^{4-n}m\int\fr{d^nq}{(2\pi)^n}\fr{\gb^{\mu\nu}}{\gb^{\aa\bb}q_\aa q_\bb-m^2+i\veps}(-iM_{\mu\nu}(q)),
\label{PROPREN1A}
\ee
and
\be
\SS^\tau(0)=-(ie)^2\mu^{4-n}\int\fr{d^nq}{(2\pi)^n}\fr{\ggb^\mu(\ggb^\ss q_\ss-m)\ggb^\tau (\ggb^\rho q_\rho-m)\ggb^\nu}{(\ggb^{\aa\bb}q_\aa q_\bb-m^2)^2}(-iM_{\mu\nu}(q)).
\label{PROPREN1B}
\ee
It follows that
\be
\SS^\tau(0)=H^\tau_{~~\rho}\ggb^\rho,
\ee
where
\be
H^\tau_{~~\rho}=H^{(1)\tau}_{~~~~\rho}+H^{(2)\tau}_{~~~~\rho},
\label{PROPREN1C}
\ee
with
\be
H^{(1)\tau}_{~~~~~\rho}=(ie)^2\mu^{4-n}\int\fr{d^nq}{(2\pi)^n}\fr{\dd^\mu_\rho\gb^{\tau\nu}+\dd^\nu_\rho\gb^{\tau\mu}-\dd^\tau_\rho\gb^{\mu\nu}}{\gb^{\aa\bb}q_\aa q_\bb-m^2+i\veps}(-iM_{\mu\nu}(q)),
\label{PROPREN1D}
\ee
and
\be
H^{(2)\tau}_{~~~~~\rho}=-2(ie)^2\mu^{4-n}\int\fr{d^nq}{(2\pi)^n}\fr{\gb^{\tau\bb}q_\bb q_\ss(\dd^\mu_\rho\gb^{\ss\nu}+\dd^\nu_\rho\gb^{\ss\mu}-\dd^\ss_\rho\gb^{\mu\nu})}{(\gb^{\aa\bb}q_\aa q_\bb-m^2+i\veps)^2}(-iM_{\mu\nu}(q)).
\label{PROPREN1E}
\ee

It is useful to split $H^\tau_{~~\rho}$ into a trace part and a traceles part,
\be
H^\tau_{~~\rho}=\fr{1}{n}H\dd^\tau_\rho + h^\tau_{~~\rho},
\ee
where
\be 
H=H^\tau_{~~\tau},
\ee
and 
\be
h^\tau_{~~\tau}=0.
\ee
The pole at $n=4$ in $H$ determines the field renormalisation of the electron propagator 
while the pole in $h^\tau_{~~\rho}$ fixes the counter term in $e^\mu_{0~a}$. We have then from
eq(\ref{PROPREN1})
\be
S^{-1}(p)=\ggb_0^\mu p_\mu-m_0+\SS(0)+\SS^\mu(0)p_\mu,
\ee
where we retain only the pole contributions in $\SS(0)$ {\it etc}. Using eq(\ref{REN2}) and eq(\ref{REN4})
we have
\be
S^{-1}(p)=(\eb^\mu_{~~a}+e^2\eb^{(1)\mu}_{~~~~~a})\gg^ap_\mu-m(1+e^2b^{(1)})
           +\SS(0)+(h^\mu_{~~\rho}+\fr{1}{4}H\dd^\mu_\rho)\ggb^\rho p_\mu.
\ee
If we set
\be
e^2\eb^{(1)\mu}_{~~~~~a}\eb^a_{~~\rho}=-h^\mu_{~~\rho};
\label{PROPREN2}
\ee
and
\be
me^2b^{(1)}=\fr{1}{4}mH+\SS(0)
\label{PROPREN3}
\ee
then eq(\ref{PROPREN1}) becomes
\be
S^{-1}(p)=(1+\fr{1}{4}H)(\ggb^\mu p_\mu-m).
\label{PROPREN4}
\ee
Finally we see that the field renormalisation for the electron is
\be
Z_e=(1-\fr{1}{4}H),
\label{PROPREN5}
\ee
Hence the renormalized inverse propagator $S^{-1}_R(p)=Z_eS^{-1}(p)$ is finite to $O(e^2)$.
It is useful to note that eq(\ref{PROPREN2}) implies 
\be
\gb^{\mu\nu}_0=\gb^{\mu\nu}-h^\mu_{~~\rho}\gb^{\rho\nu}-h^\nu_{~~\rho}\gb^{\rho\mu}.
\label{PROPREN6}
\ee

\section{\label{VTX} Vertex}

\begin{figure}[t]
   \centering
  \includegraphics[width=0.6\linewidth]{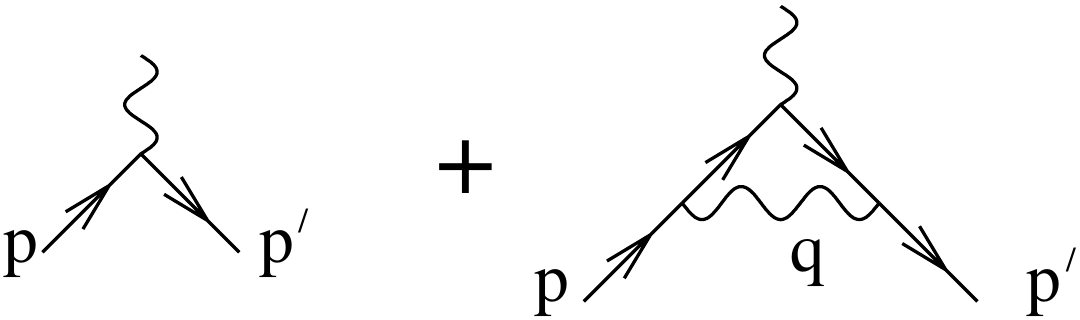}
  \caption{Contributions to the electron-photon vertex to $O(e_0^3)$. }
  \label{FIG3}
\end{figure}

The complete vertex amplitude to $O(e^3)$ corresponds to the diagrams in Fig \ref{FIG3}.
It has the form $V^\mu(p,p')=ie_0\ggb^\tau+\Vc^\tau(p,p')$ where

\be
\Vc^\tau(p,p')=(ie)^3\int\fr{d^nq}{(2\pi)^n}\fr{\ggb^\mu i(\ggb^\ss(p'_\ss-q_\ss)+m)\ggb^\tau i(\ggb^\rho(p_\rho-q_\rho)+m)\ggb^\nu(-iM_{\mu\nu}(q))}{(\gb^{\aa'\bb'}(p'_{\aa'}-q_{\aa'})(p'_{\bb'}-q_{\bb'})-m^2)(\gb^{\aa\bb}(p_\aa-q_\aa)(p_\bb-q_\bb)-m^2)}.
\ee
The divergence is contained in
\be
\Vc^\tau(0,0)=-(ie)^3\int\fr{d^nq}{(2\pi)^n}\ggb^\mu\fr{(\gb^\ss q_\ss-m)\ggb^\tau(\ggb^\rho q_\rho-m)}{(\ggb^{\aa\bb}q_\aa q_\bb-m^2)^2}\ggb^\nu(-iM_{\mu\nu}(q)).
\ee
It follows that 
\be
\Vc^\tau(0,0)=ie\SS^\tau(0).
\ee
Hence the divergence of $\Vc^\tau(0,0)$ is the same as that of $\SS^\tau(0)$ up to a factor $ie$.
Using eq(\ref{REN1}) we have
\be
V^\mu(0,0)=\mu^{4-n}ie(1+e^2a^{(1)})(\eb^\mu_{~~a}+e^2\eb^{(1)\mu}_{~~~~~a})\gg^a+ie\SS^\mu(0).
\ee
It follows that if we use eq(\ref{PROPREN2}) we find to $O(e^3)$
\be
V^\mu(0,0)=ie((1+\fr{1}{4}H+e^2a^{(1)})\ggb^\mu=ie(1+\fr{1}{4}H)(1+e^2a^{(1)})\ggb^\mu.
\ee
The renormalized vertex $V_R^\mu(0,0)=Z_e\sqrt{Z_\gg}V^\mu(0,0)$ is finite provided we set
\be
1+e^2a^{(1)}=\fr{1}{\sqrt{Z_\gg}},
\ee
that is
\be
e^2a^{(1)}=-\fr{e^2}{144\pi^2}\fr{W}{n-4},
\label{CHARGEREN1}
\ee
or
\be
e_0^2=\mu^{4-n}e^2\left(1-\fr{e^2}{72\pi^2}\fr{W}{n-4}\right).
\label{CHARGEREN2}
\ee

\section{\label{POLES} Pole Divergences at $n=4$}

The pole divergences at $n=4$ can be exhibited explicitly by making use of
the photon propagator representation in eq(\ref{INVREP}). For example
from eq(\ref{PROPREN1A}) we find
\be
\SS(0)=-i(ie)^2\mu^{4-n}m\gb^{\mu\nu}g_{\mu\rho}\exp\{-i\Nbf(\d_z)\}^\rho_{~~\nu}L(z),
\label{DIV1}
\ee
where
\be
L(z)=-i\int du\int\fr{d^nq}{(2\pi)^n}\fr{1}{\gb^{\aa\bb}q_\aa q_\bb-m^2+i\veps}e^{iu(q^2+z.q/\sqrt{u}+i\veps)},
\ee
and where finally we set $z=0$. We can now express $L(z)$ in the form
\begin{eqnarray}
L(z)&=&(-i)^2\int dudv\int\fr{d^nq}{(2\pi)^n}e^{iu(q^2+z.q/\sqrt{u})+iv(\gb^{\aa\bb}q_\aa q_\bb-m^2+i\veps)}\nonumber\\
    &=&(-i)^2\int_0^1dx\int_0^\infty d\ll\ll\int\fr{d^nq}{(2\pi)^n}e^{i\ll(\ght^{\aa\bb}(x)q_\aa q_\bb+xz^\aa q_\aa/\sqrt{u}-m^2(1-x)+i\veps)},
\end{eqnarray}
where we have set $u=\ll x$ and $v=\ll(1-x)$ and have introduced the 
interpolated (inverse) metric $\ght^{\aa\bb}(x)=xg^{\aa\bb}+(1-x)\gb^{\aa\bb}$. This metric was introduced in reference \cite{ITD2}
where it was emphasised that it should remain non-singular for $0\le x\le 1$ as a condition of 
acceptable causal structure. This was ensured by the requirement that lightcones associated 
with $g_{\aa\bb}$ and $\gb_{\aa\bb}$ overlap so that there exists a shared set of spacetime vectors 
that are timelike in both metrics. The same point holds here. Making this assumption we can evaluate $L(z)$ as
\be
L(z)=(-i)^2\int_0^1dx\int_0^\infty d\ll\ll\int\fr{d^nq'}{(2\pi)^n}e^{i\ll(\ght^{\aa\bb}q'_\aa q'_\bb+x^2\ght_{\aa\bb}z^\aa z^\bb/(4u)-m^2(1-x)+i\veps)},
\ee
where $q_\aa=q'_\aa-x\ght_{\aa\bb}z^\bb$ and $\ght_{\aa\bb}(x)$ is the inverse of $\ght^{\aa\bb}(x)$. 
The non-singularity condition on $\ght^{\aa\bb}(x)$ allows us to evaluate the gaussian integral 
yielding 
\be
L(z)=(-i)^2\int_0^1dx\int_0^\infty d\ll\ll i\left(\fr{\pi}{i\ll}\right)^{n/2}\fr{1}{\sqrt{-\det\ght^{\aa\bb}(x)}}e^{x\ght_{\aa\bb}(x)z^\aa z^\bb/4}e^{-i\ll m^2(1-x)}.
\ee
Hence
\begin{eqnarray}
L(z)&=&\fr{i}{(4\pi)^{n/2}}\int_0^1dx\GG(2-\fr{n}{2})(m^2(1-x))^{n/2-2}\fr{1}{\sqrt{-\det\ght^{\aa\bb}(x)}}e^{ix\ght^{\aa\bb}(x)z^\aa z^\bb/4}\nonumber\\
  &=&-\fr{i}{8\pi^2}\fr{1}{n-4}\int_0^1dx\fr{1}{\sqrt{-\det\ght^{\aa\bb}(x)}}e^{i\ght_{\aa\bb}(x)z^\aa z^\bb/4}.
\label{POL1}
\end{eqnarray}
Finally we can reconstruct $\SS(0)$ using eq(\ref{DIV1}).

Similarly we see that
\be
H^{(1)\tau}_{~~~~\rho}=ie^2\mu^{4-n}(\dd^\mu_\rho \gb^{\tau\nu}+\dd^\nu_\rho\gb^{\tau\mu}-\dd^\tau_\rho\gb^{\mu\nu})g_{\mu\ll}(e^{-i\Nbf(-i\d_z)})^\ll_{~~\nu}L(z).
\ee
From a slightly more complex calculation we find
\be
H^{(2)\tau}_{~~~~\rho}=-2ie^2\gb^{\tau\ll}(\dd^\mu_\rho\gb^{\ss\nu}+\dd^\nu_\rho\gb^{\ss\mu}-\dd^\ss_\rho\gb^{\mu\nu})g_{\mu\kk}(e^{-i\Nbf(-i\d_z)})^\kk_{~~\nu}L_{\ll\ss}(z).
\ee
where
\be
L_{\ll\ss}(z)=-i\int du\int\fr{d^nq}{(2\pi)^n}\fr{q_\ll q_\ss}{(\gb^{\aa\bb}q_\aa q_\bb-m^2)^2}e^{iu(q^2+z.q/\sqrt{u}+i\veps)}.
\ee
The pole at $n=4$ has the form
\be
L_{\ll\ss}(z)=-\fr{1}{8\pi^2}\fr{1}{n-4}\int_0^1dx\fr{(1-x)}{\sqrt{-\det\ght^{\aa\bb}(x)}}(\fr{i}{2}\ght_{\ll\ss}(x)+\fr{x}{4}z^\aa z^\bb\ght_{\ll\aa}(x)\ght_{\ss\bb}(x))e^{ix\ght_{\aa\bb}(x)z^\aa z^\bb/4}.
\label{POL2}
\ee

\section{\label{WLTBM} Petrov Class of Bimetrically Generated WLTs}

It would be interesting to develop a general theory of how the bimetric structure of the theory
affects the nature of Lorentz symmetry breaking through the renormalisation process.
At present this seems rather difficult. For now we confine attention to some particular
examples involving the simpler Petrov classes.
We will adopt a minimal approach that assumes presence of contributions to the WLT $C^{\mu\nu\ss\tau}$
only of a type forced on us by the need to accomodate the divergences accompanying the WLT $V^{\mu\nu\ss\tau}$
in section \ref{VACP}. 

\subsection{\label{WLT_0} Class O}

In fact the simplest non-trivial case is class O for which $V^{\mu\nu\ss\tau}=0$. There are three
cases. They arise
when there is a coordinate frame for which $\gb^{\mu\nu}$
has the form
\be
\gb^{\mu\nu}=bg^{\mu\nu}\pm(a-b)k^\mu k^\nu.
\label{METRIC_O}
\ee
We choose $+$ or $-$ according as $k^\mu$ is timelike $(g_{\mu\nu} k^\mu k^\nu>0)$ or spacelike  $(g_{\mu\nu} k^\mu k^\nu<0)$.
In order to maintain $\det \gb^{\mu\nu}=-1$ we impose $ab^3=1$. 
There is a third case $k^\mu=l^\mu$ where the lightlike vector $l^\mu$ satisfies $g_{\mu\nu}l^\mu l^\nu=0$
and 
\be
\gb^{\mu\nu}=g^{\mu\nu}+wl^\mu l^\nu.
\ee
It is easy to verify that the WLT $V^{\mu\nu\ss\tau}$ vanishes in all three cases. In our minimal approach we therefore
assume $C^{\mu\nu\ss\tau}$ vanishes. In the timelike case the underlying reason, of course, is that we are maintaining 
invariance under the rotation group that leaves $k^\mu$ invariant and a WLT cannot exhibit such an invariance unless it
is null. Similar remarks apply in the other cases. Under these circumstances although we still have Lorentz symmetry 
breaking, the lightcone associated with photons being distinct from that associated with electrons, we do not have 
birefringence for the photons.

\subsection{\label{WLT_N} Class N}

The next most simple case is class N. Such a Lorentz symmetry breaking situation may be induced
by using the NP tetrad to express the electron inverse metric in the form
\be
\gb^{\mu\nu}=g^{\mu\nu}+l^\mu(\cb m^\nu+c\mb^\nu)+(\cb m^\mu+c\mb^\mu)l^\nu+\aa l^\mu l^\nu,
\label{METN1}
\ee
where for the moment $\aa$ is an arbitrary real parameter.
It is readily verified that in this case $\det \gb^{\mu\nu}=-1$ and
\be
V^{\mu\nu\ss\tau}=\cb^2A^{\mu\nu}A^{\ss\tau}+c^2\Ab^{\mu\nu}\Ab^{\ss\tau}.
\label{METN2}
\ee
These two cases will be examined in detail later. However a further example provides
additional insight into the effect of vacuum polarisation in the bimetric context.

\subsection{\label{WLT_D} Class D}

One simple way of constructing a new metric from the standard one is to consider a
shearing of space-time. Such a shearing is represented by the mapping $x^\mu\rightarrow x'^\mu=T^\mu_{~~\nu}x^\nu$,
where
\be
T^\mu_{~~\nu}=\dd^\mu_\nu+vf^\mu h_\ss,
\ee
and
\begin{eqnarray}
h^2&=&1\nonumber\\
f^2&=&-1\nonumber\\
f.h&=&0.
\end{eqnarray}
We can then define a new metric to be of the form
\be
\gb^{\mu\nu}=T^\mu_{~~\ss}T^\nu_{~~\tau}\eta^{\ss\tau}.
\ee
If now we calculate $W^{\mu\nu\ss\tau}=\ggb^{\mu\ss}\ggb^{\nu\tau}-\ggb^{\mu\tau}\ggb^{\nu\ss}$,
we obtain
\begin{eqnarray}
W^{\mu\nu\ss\tau}&=&\eta^{\mu\ss}\eta^{\nu\tau}-\eta^{\mu\tau}\eta^{\nu\ss}\nonumber\\
                 &&+\eta^{\mu\ss}(v(f^\nu h^\tau+f^\tau h^\nu)+v^2f^\nu f^\tau)\nonumber\\
                 &&+\eta^{\nu\tau}(v(f^\mu h^\ss+f^\ss h^\mu)+v^2f^\mu f^\ss)\nonumber\\
                 &&-\eta^{\mu\tau}(v(f^\nu h^\ss+f^\ss h^\nu)+v^2f^\nu f^\ss)\nonumber\\
                 &&-\eta^{\nu\ss}(v(f^\mu h^\tau+f^\tau h^\mu)+v^2f^\mu f^\tau)\nonumber\\
                 &&-v^2(f^\mu h^\nu-f^\nu h^\mu)(f^\ss h^\tau-f^\tau h^\ss).
\end{eqnarray}
When we extract $V^{\mu\nu\ss\tau}$ we obtain
\begin{eqnarray}
V^{\mu\nu\ss\tau}&&-v^2\{\fr{1}{3}(\eta^{\mu\ss}\eta^{\nu\tau}-\eta^{\mu\tau}\eta^{\nu\ss})\nonumber\\
                 &&+\fr{1}{2}\eta^{\mu\ss}(f^\nu f^\tau-h^\nu h^\tau)\nonumber\\
                 &&+\fr{1}{2}\eta^{\nu\tau}(f^\mu f^\ss-h^\mu h^\ss)\nonumber\\
                 &&-\fr{1}{2}\eta^{\mu\tau}(f^\nu f^\ss-h^\nu h^\ss)\nonumber\\
                 &&-\fr{1}{2}\eta^{\nu\ss}(f^\mu f^\tau-h^\mu h^\tau)\nonumber\\
                 &&-(f^\mu h^\nu-f^\nu h^\mu)(f^\ss h^\tau-f^\tau h^\ss)\}.
\end{eqnarray}
Setting $f^\mu=(l^\mu+n^\mu)/\sqrt{2}$ and $h^\mu=(l^\mu-n^\mu)/\sqrt{2}$ we find
\be
V^{\mu\nu\ss\tau}=\fr{1}{6}v^2\{A^{\mu\nu}B^{\ss\tau}+B^{\mu\nu}A^{\ss\tau}+D^{\mu\nu}D^{\ss\tau}\}+\mbox{c.c.}.
\ee
In this case of sheared lightcones we find that the WLT is indeed of Petrov class D. However since the coefficient
is real it is not quite the most general case for class D.

The remaining classes, although obviously worth investigating, are considerably more elaborate.
For example the case for which $\gb^{\mu\nu}$ is obtained from $g^{\mu\nu}$ by separate rescalings 
in each of the timelike and spacelike directions leads to a WLT of Petrov class I. We postpone such 
a completion of our program for a later discussion.

\section{\label{RENEX} Special Examples of Renormalized Bimetric Theory.}

We look in more detail at renormalisation in two special cases, Petrov classes O and N that
are particularly tractable.

\subsection{\label{RENEX0} Bimetric Theory with Petrov Class O }

It is convenient to construct the the three metric cases in class O by introducing a reference metric 
which we choose to be the standard Lorentz metric $\eta^{\mu\nu}$. For the timelike case
we have
\begin{eqnarray}
g_0^{\mu\nu}&=&\bb_0\eta^{\mu\nu}+(\aa_0-\bb_0) k^\mu k^\nu,\nonumber\\
g^{\mu\nu}&=&\bb\eta^{\mu\nu}+(\aa-\bb) k^\mu k^\nu,\nonumber\\
\gb_0^{\mu\nu}&=&\bbb_0\eta^{\mu\nu}+(\aab_0-\bbb_0) k^\mu k^\nu,\nonumber\\
\gb^{\mu\nu}&=&\bbb\eta^{\mu\nu}+(\aab-\bbb) k^\mu k^\nu,
\end{eqnarray}
where $k^\mu=(1,0,0,0)$ and hence $\eta_{\mu\nu}k^\mu k^\nu=1$. We also find it convenient to set $\aab=a\aa$ and $\bbb=b\bb$.
Each of the above metrics has a determinant of $-1$. In particular we have $\aa\bb^3=\aab\bbb^3=1$. 
This implies that $ab^3=1$. The inverses of the above metrics are easily constructed by 
the replacements $\aa\rightarrow\aa^{-1}$ and $\bb\rightarrow\bb^{-1}$ {\it etc}. We have also
\be
\gb^{\mu\nu}=bg^{\mu\nu}+\aa(a-b)k^\mu k^\nu.
\ee
We then find that
\be
W^{\mu\nu\ss\tau}=b^2(g^{\mu\ss}g^{\nu\tau}-g^{\mu\tau}g^{\nu\ss}) +\aa(a-b)(g^{\mu\ss}k^\nu k^\tau
                         +g^{\nu\tau}k^\mu k^\ss-g^{\mu\tau}k^\nu k^\ss-g^{\nu\ss}k^\mu k^\tau)
\ee
It follows that
\be
V^{\mu\ss}=\fr{1}{2}b(b-a)(g^{\mu\ss}-4\aa k^\mu k^\ss),
\ee
and
\be
W=6b(a+b)=6(b^2+b^{-2}).
\ee
It is easily confirmed that the WLT $V^{\mu\nu\ss\tau}$ vanishes. The pole divergence in $C_0^{\mu\nu\ss\tau}$ 
at $n=4$ therefore also vanishes and we are free to apply our minimal assumption that $C^{\mu\nu\ss\tau}=0$.
In that case the representation of the photon propagator simplifies and we can deduce
\be
\SS(0)=-i(ie)^2\mu^{4-n}m\gb^{\mu\nu}g_{\mu\nu}L(0)=m\fr{e^2}{8\pi^2}\fr{1}{n-4}\gb^{\mu\nu}g_{\mu\nu}\int_0^1dx\fr{1}{\sqrt{-\det\ght^{\aa\bb}(x)}},
\ee
and
\begin{eqnarray}
H^{(1)\tau}_{~~~~\rho}&=&-i(ie)^2\mu^{4-n}(2g_{\mu\rho}\gb^{\tau\mu}-\dd^\tau_\rho \gb^{\mu\nu}g_{\mu\nu})L(0)\nonumber\\
  &=&\fr{e^2}{8\pi^2}(2g_{\mu\rho}\gb^{\tau\mu}-\dd^\tau_\rho \gb^{\mu\nu}g_{\mu\nu})\fr{1}{n-4}\int_0^1dx\fr{1}{\sqrt{-\det\ght^{\aa\bb}(x)}}.
\end{eqnarray}
From eq(\ref{POL2}) we have
\begin{eqnarray}
H^{(2)\tau}_{~~~~\rho}&=&2i(ie)^2\mu^{4-n}(2g_{\rho\nu}\gb^{\ss\nu}-\dd^\ss_\rho g_{\mu\nu}\gb^{\mu\nu})\gb^{\tau\bb}L_{\bb\ss}(0)\nonumber\\
    &=&-\fr{e^2}{8\pi^2}\fr{1}{n-4}(2g_{\rho\nu}\gb^{\ss\nu}-\dd^\ss_\rho \gb^{\mu\nu}g_{\mu\nu})\gb^{\tau\bb}\int_0^1dx\fr{1-x}{\sqrt{-\det\ght^{\aa\bb}(x)}}\ght_{\bb\ss}(x).
\end{eqnarray}
Using the explicit form for $g^{\mu\nu}$ and $k^\mu$ indicated in subsection \ref{WLT_0} we have
\be
\gb^{\aa\bb}=\left(
\begin{array}{cccc} 
a\aa&0&0&0\\
0&-b\bb&0&0\\
0&0&-b\bb&0\\
0&0&0&-b\bb
\end{array}
\right)
\ee
We have then
\be
\ght^{\aa\bb}(x)=\left(
\begin{array}{cccc}
\aa(a+(1-a)x)&0&0&0\\
0&-\bb((b+(1-b)x))&0&0\\
0&0&-\bb(b+(1-b)x))&0\\
0&0&0&-\bb(b+(1-b)x))
\end{array}
\right)
\ee
The inverse matrix $\ght_{\aa\bb}(x)$ is obvious and 
\be
-\det \ght^{\aa\bb}(x)=(a+(1-a)x)(b+(1-b)x)^3.
\ee
We require the integrals
\be
J_0=\int_0^1dx\fr{1}{(a+(1-a)x)^{1/2}(b+(1-b)x)^{3/2}}=\fr{2}{\sqrt{b}(\sqrt{a}+\sqrt{b})}.
\ee
\be
J_1=\int_0^1dx\fr{1-x}{(a+(1-a)x)^{3/2}(b+(1-b)x)^{3/2}}=\fr{2}{\sqrt{ab}(\sqrt{a}+\sqrt{b})^2}.
\ee
\be
J_2=\int_0^1dx\fr{1-x}{(a+(1-a)x)^{1/2}(b+(1-b)x)^{5/2}}=\fr{2}{3}\fr{(\sqrt{a}+2\sqrt{b})}{b^{3/2}(\sqrt{a}+\sqrt{b})^2}.
\ee
The poles at $n=4$ are then
\be
\SS(0)=m\fr{e^2}{8\pi^2}\fr{2(a+3b)}{\sqrt{b}(\sqrt{a}+\sqrt{b})}\fr{1}{n-4}.
\ee
\be
H^{(1)\tau}_{~~~~\rho}=\fr{e^2}{8\pi^2}
\left(
\begin{array}{cccc}
(a-3b)J_0&0&0&0\\
0&-(a+b)J_0&0&0\\
0&0&-(a+b)J_0&0\\
0&0&0&-(a+b)J_0
\end{array}
\right)\fr{1}{n-4}.
\ee
\be
H^{(2)\tau}_{~~~~\rho}=-\fr{e^2}{8\pi^2}\left(
\begin{array}{cccc}
a(a-3b)J_1&0&0&0\\
0&-b(a+b)J_2&0&0\\
0&0&-b(a+b)J_2&0\\
0&0&0&-b(a+b)J_2
\end{array}
\right)\fr{1}{n-4}.
\ee
\be
H^{(1)}=H^{(1)\tau}_{~~~~\tau}=-\fr{e^2}{8\pi^2}2(a+3b)J_0\fr{1}{n-4}.
\ee
The traceless part is
\be
h^{(1)\tau}_{~~~~\rho}=H^{(1)\tau}_{~~~~\rho}-\fr{1}{4}H^{(1)}\dd^\tau_\rho
      =\fr{e^2}{8\pi^2}\fr{3}{2}(a-b)J_0T^\tau_{~~\rho}\fr{1}{n-4},
\ee
where $T^\tau_{~~\rho}$ is the diagonal traceless matrix with diagonal
entries $(1,-1/3,-1/3,-1/3)$. We have also
\be
H^{(2)}=H^{(2)\tau}_{~~~~\tau}=\fr{e^2}{8\pi^2}\fr{4}{n-4}.
\ee
The traceless part is
\be
h^{(2)\tau}_{~~~~\rho}
  =-\fr{e^2}{8\pi^2}\fr{2a^{3/2}+a\sqrt{b}-4\sqrt{a}b+b^{3/2}}{\sqrt{b}(\sqrt{a}+\sqrt{b})^2}T^\tau_{~~\rho}\fr{1}{n-4}.
\ee
On combining these results and setting $a=b^{-3}$ we find
\be
H=H^{(1)}+H^{(2)}=-\fr{e^2}{8\pi^2}\fr{4(2b^4-b^2+1)}{b^2(1+b^2)}\fr{1}{n-4},
\ee
and
\be
h^\tau_{~~\rho}=h^{(1)\tau}_{~~~~\rho}+h^{(2)\tau}_{~~~~\rho}=\fr{e^2}{8\pi^2}f(b)T^\tau_{~~\rho}\fr{1}{n-4},
\ee
where
\be
f(b)=\fr{(1-b^2)(1+3b^2+4b^4)}{b^2(1+b^2)^2}.
\ee
Note that $h^\tau_{~~\rho}$ vanishes when $b=1$ as it should since this value corresponds to 
the restoration of Lorentz symmetry.

\subsubsection{\label{REGP_0}Renormalisation Group for Bimetric Theory - Petrov Class O}

With the above information we can calculate the renormalisation counterterms for the bare parameters of the theory. 
In the present model $W=6(b^2+b^{-2})$ with the result that
\be
e_0^2=\mu^{4-n}e^2\left(1-\fr{e^2}{12\pi^2}\fr{(b^2+b^{-2})}{n-4}\right).
\label{RENEX1}
\ee
Assuming that the renormalisation process works beyond our second order calculation
we can explore the implication of the renormalisation group for this model.
Setting $t=\log(\mu/\mu_S)$ where $\mu_S$ is a standard scale for which the corresponding renormalised 
charge, $e_S$ is small,
we can use the lack of dependence of the bare parameter $e_0$ on $\mu$ to deduce that
\be
\fr{d}{dt}e_0^2=0.
\ee
It follows then from eq(\ref{RENEX1}) to $O(e^2)$ that
\be
\fr{d}{dt}(e^2)=e^2\left(-(4-n)+\fr{e^2}{12\pi^2}(b^2+b^{-2})\right)
\label{REGP_01}
\ee
The bare metric $g^{\mu\nu}_0$ is a diagonal matrix with entries $(\aa_0,\bb_0,\bb_0,\bb_0)$. 
We can infer to $O(e^2)$ from eq(\ref{PROPREN6}) that
\be
e^2g^{(1)\mu\nu}=\fr{e^2}{24\pi^2}b(b-a)\fr{1}{n-4}(\bb\eta^{\mu\nu}-(\bb+3\aa)k^\mu k^\nu)
\label{RENEX2}
\ee
We find then
\begin{eqnarray}
\aa_0&=&\aa\left(1-3\fr{e^2}{24\pi^2}b(b-a)\fr{1}{n-4}\right),\nonumber\\
\bb_0&=&\bb\left(1+\fr{e^2}{24\pi^2}b(b-a)\fr{1}{n-4}\right).
\end{eqnarray}
Note that these results are of course consistent with (to $O(e^2)$) with the relation $\aa_0\bb_0^3=1$.
Again the bare parameter $\bb_0$ is independent of $\mu$ therefore we can conclude that
\be
\fr{d\bb}{dt}=-\bb\fr{e^2}{24\pi^2}b(b-a)=-\bb\fr{e^2}{24\pi^2}(b^2-b^{-2}).
\ee

From eq(\ref{PROPREN6}) we can deduce that
\begin{eqnarray}
\aab_0&=&\aab\left(1-\fr{e^2}{4\pi^2}f(b)\fr{1}{n-4}\right),\nonumber\\
\bbb_0&=&\bbb\left(1+\fr{e^2}{12\pi^2}f(b)\fr{1}{n-4}\right).
\end{eqnarray}
This is consistent with $\aab_0\bbb_0^3=1$, and we have
\be
\fr{d\bbb}{dt}=-\bbb\fr{e^2}{12\pi^2}f(b).
\ee

Recalling $\bbb=b\bb$ we have
\be
\fr{\bbb_0}{\bb_0}=b\left(1+\fr{e^2}{24\pi^2}F(b)\fr{1}{n-4}\right),
\ee
where
\be
F(b)=2f(b)-(b^2-b^{-2}).
\ee
We obtain the result
\be
\fr{db}{dt}=-b\fr{e^2}{24\pi^2}F(b).
\ee
From eq(\ref{PROPREN4}) we see that
\be
m_0=m\left(1+\fr{e^2}{8\pi^2}\fr{4b^4+b^2+1}{b^2(1+b^2)}\fr{1}{n-4}\right).
\ee
The renormalisation group equation is
\be
\fr{dm}{dt}=-\fr{e^2}{8\pi^2}\fr{4b^4+b^2+1}{b^2(1+b^2)}m.
\label{REGP_02}
\ee

These RG equations have a particularly significant  fixed point at $e^2=0$ and $b=1$ which corresponds
to the Lorentz invariant case at zero coupling. For small departures from Lorentz invariance, $b=1+y$
where $y$ is small, we find on expanding in powers of $y$ and retaining only linear terms
\be
\fr{d(e^2)}{dt}=\fr{e^4}{6\pi^2},
\label{REGP_04}
\ee
and
\be
\fr{dy}{dt}=3\fr{e^2}{6\pi^2}y.
\label{REGP_05}
\ee
The solution for the RG trajectory in the neighbourhood of the fixed point is
\be
\fr{e^2}{e_S^2}=\left(1-\fr{e_S^2}{6\pi^2}t\right)^{-1},
\ee
and
\be
\fr{y}{y_S}=\left(1-\fr{e_S^2}{6\pi^2}t\right)^{-3},
\ee
where $e_S$ and $y_S$ are the assigned values of $e$ and $y$ at $t=0$ or $\mu=\mu_S$.
This shows that the fixed point is IR attractive and that in its neighbourhood we have
\be
\fr{y}{y_S}=\left(\fr{e^2}{e_S^2}\right)^3.
\ee
Here $e^2_S$ and $y_S$ are the coupling and (small) departure from Lorentz invariance at 
the standard scale $\mu=\mu_S$. It follows that in the IR limit the theory exhibits the same
behaviour as as implied by the analysis in references \cite{NLSN1,KOST4}. In the same approximation 
we find from eq(\ref{REGP_02})
\be
\fr{dm}{dt}=-\fr{3e^2}{8\pi^2}m.
\ee
and therefore we have the result, identical with that for the Lorentz case,
\be
m=m_S\left(1-\fr{e_S^2}{6\pi^2}t\right)^{9/4},
\ee
where $m_S$ is the value of the mass parameter when  $\mu=\mu_S$.

\begin{figure}[t]
   \centering
  \includegraphics[width=0.6\linewidth]{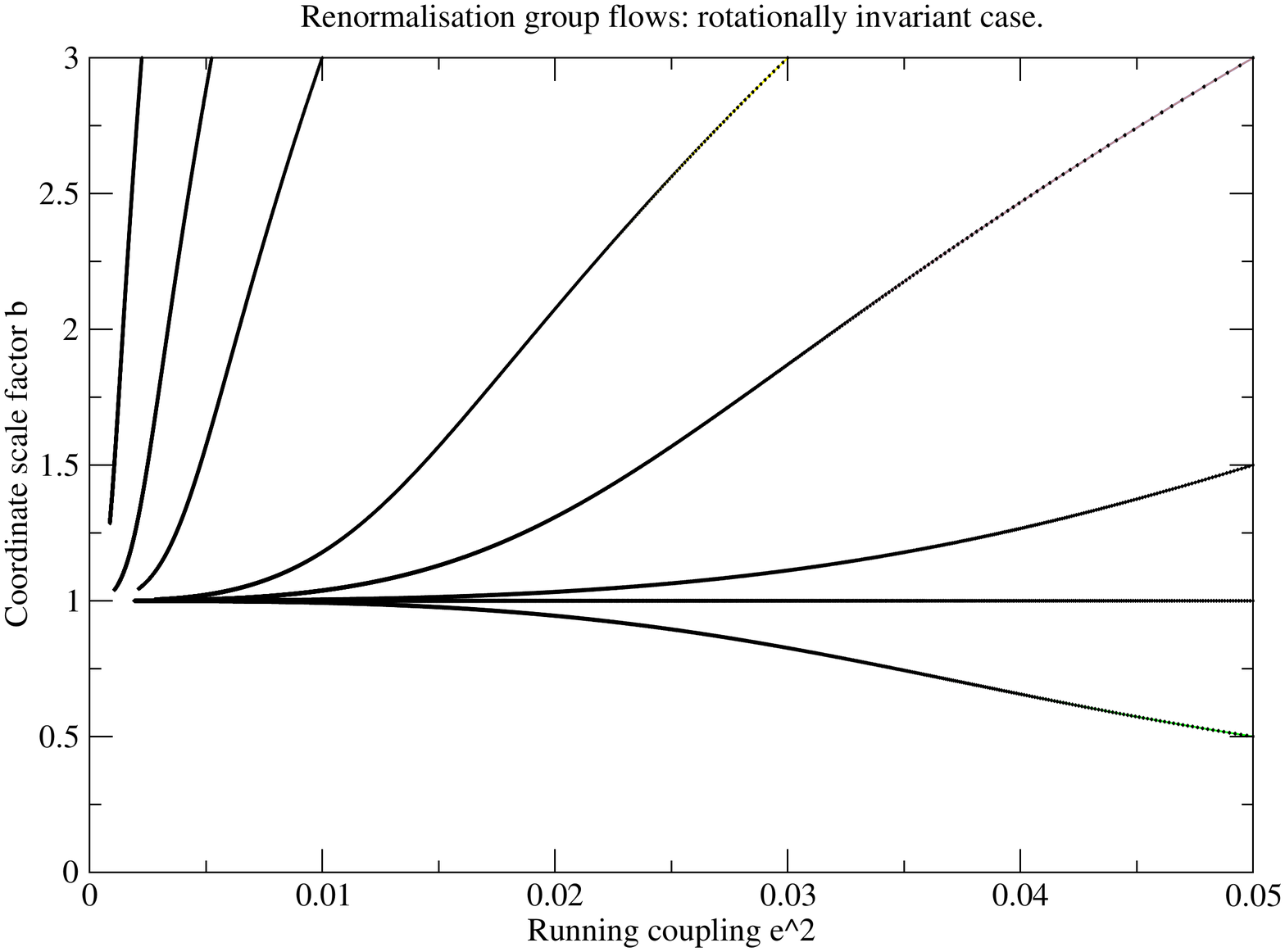}
  \caption{Renormalisation group trajectories Petrov class O: rotationally invariant case. }
  \label{FIG4}
\end{figure}

This gives a description of the behaviour of the effective parameters in the neighbourhood of the 
point $e^2=0$, $b=1$.
However in our approach we can compute the complete RG trajectory without constraint on $b$,
provided of course that $e^2$ does not become too large. The results are illustrated in Fig.\ref{FIG4}.
An important observation is that no matter how small $e^2$ or how large $b$ the RG trajectory 
never approaches the axis $e^2=0$ except at the fixed point discussed above. 
The axis $e^2=0$ is of course a line of fixed points corresponding to 
a theory with no coupling between electrons and photons. Such a non-interacting theory can 
maintain any Lorentz symmetry breaking imposed on it. The conclusion is then that no matter how weak
the electron-photon coupling or how large the Lorentz symmetry breaking at higher energies
the theory will at least in the massless case exhibit Lorentz symmetry in the IR limit as proposed in
reference \cite{NLSN1}.

The case in which the vector $k^\mu$ is spacelike results in an essentially identical analysis
which need not be repeated explicitly. The case in which $k^\mu$ is replaced by a lightlike vector is 
different in detail but yields essentially similar results, in particular that Lorentz symmetry is
restored in the IR limit. It can be viewed as a special case of the model 
discussed in the next section and we do not treat it separately.

\subsection{\label{RENEXN} Bimetric Theory with Petrov Class N }

The next most tractable example is provided by the choice that $\gb^{\mu\nu}$ has the structure
exhibited in eq(\ref{METN1}). There is no loss of generality in choosing the parameter $c$ to be real.
In order to parametrise the various metrics we introduce in this case a reference metric which
we are free to choose to be $\eta^{\mu\nu}$
and an associated NP tetrad $l,n,m,\mb$ with the properties $l^\mu=\eta^{\mu\nu}l_\nu$, $l^2=n^2=m^2=\mb^2=0$,
$l.n=-m.\mb=1$, and $l.m=l.\mb=n.m=n.\mb=0$. We now construct the parametrised metrics in the form
\begin{eqnarray}
g_0^{\mu\nu}&=&\eta^{\mu\nu}+s_0P^{\mu\nu}+r_0l^\mu l^\nu,\nonumber\\
g^{\mu\nu}&=&\eta^{\mu\nu}+sP^{\mu\nu}+rl^\mu l^\nu,\nonumber\\
\gb_0^{\mu\nu}&=&\eta^{\mu\nu}+u_0P^{\mu\nu}+v_0l^\mu l^\nu,\nonumber\\
\gb^{\mu\nu}&=&\eta^{\mu\nu}+uP^{\mu\nu}+vl^\mu l^\nu,
\end{eqnarray}
where
\be
P^{\mu\nu}=l^\mu(m^\nu+\mb^\nu)+(m^\mu+\mb^\mu)l^\nu.
\ee
Of course the parameters $s,r,u,v$ are the renormalised versions of $s_0,r_0,u_0,v_0$ 
so that $s_0=s+e^2s^{(1)}$ to $O(e^2)$ where $s^{(1)}$ has  a pole at $n=4$ and similarly for the other parameters.
We also require an NP tetrad $l(s),m(s),n(s),\mb(s)$ associated with the metric $g^{\mu\nu}$. 
We achieve this by setting $l_\mu(s)=l_\mu$, $m_\mu(s)=m_\mu$. By imposing the relations
$l^\mu(s)=g^{\mu\nu}l_\nu(s)$ etc, we find
\begin{eqnarray}
l^\mu(s)&=&l^\mu\nonumber,\\
m^\mu(s)&=&m^\mu-sl^\mu\nonumber,\\
\mb^\mu(s)&=&\mb^\mu-sl^\mu.
\end{eqnarray}
It is easily checked that $l^\mu(s)l_\mu(s)=m^\mu(s)m_\mu(s)=\mb^\mu(s)\mb_\mu(s)=0$
and $l^\mu(s)m_\mu(s)=l^\mu(s)\mb_\mu(s)=0$. In addition $m^\mu(s)\mb_\mu(s)=-1$. 
Although we will not make use of it we give for completeness the form of the remaining element 
of the tetrad thus $n_\mu(s)=n_\mu-(s^2+s+r/2)l_\mu-s(m_\mu+\mb_\mu)$ and 
$n^\mu(s)=g^{\mu\nu}n_\nu(s)=n^\mu+(s^2-s+r/2)l^\mu$.
It is also easily shown that
\be
g_{\mu\nu}=\eta_{\mu\nu}-sP_{\mu\nu}-(2s^2+r)l_\mu l_\nu.
\ee
The relation between $g^{\mu\nu}$ and $\gb^{\mu\nu}$ is
\be
\gb^{\mu\nu}-g^{\mu\nu}=cP^{\mu\nu}+wl^\mu l^\nu,
\ee
where $c=u-s$ and $w=v-r$. If we define $P^{\mu\nu}(s)=l^\mu(m^\nu(s)+\mb^\nu(s))+(m^\mu(s)+\mb^\mu(s))l^\nu$
then we have
\be
P^{\mu\nu}(s)=P^{\mu\nu}-4sl^\mu l^\nu.
\ee
Hence
\be
\gb^{\mu\nu}=g^{\mu\nu}+cP^{\mu\nu}(s)+(w+4sc)l^\mu l^\nu.
\label{METN3}
\ee
This is of the same form as eq(\ref{METN1}) with the parameter $\aa=w+4sc$. The NP tetrad
is of course that appropriate to $g^{\mu\nu}$ as constructed here. The result after some algebra is that
\be
V^{\mu\nu\ss\tau}=c^2(A^{\mu\nu}A^{\ss\tau}+\Ab^{\mu\nu}\Ab^{\ss\tau}),
\ee
with $c$ real. Note that strictly speaking we should have used 
$A^{\mu\nu}(s)=l^\mu(m^\nu(s)+\mb^\nu(s))-l^\nu(m^\mu(s)+\mb^\mu(s))$ but it is 
obvious that $A^{\mu\nu}(s)=A^{\mu\nu}$. We also have $W=12$ and 
\be
V^{\mu\ss}=2cP^{\mu\ss}(s)+2(w+2sc+c^2)l^\mu l^\ss.
\ee
This may also be expressed in the form
\be
V^{\mu\ss}=2cP^{\mu\ss}+2(w-2sc+c^2)l^\mu l^\ss.
\ee
From eqs(\ref{VACP10}) and eq(\ref{CHARGEREN2}) we find here that
\begin{eqnarray}
e_0^2&=&\mu^{4-n}e^2\left(1-\fr{e^2}{6\pi^2}\fr{1}{n-4}\right),\nonumber\\
s_0&=&s+\fr{e^2}{6\pi^2}\fr{c}{n-4},\nonumber\\
r_0&=&r+\fr{e^2}{6\pi^2}\fr{(w+c^2)}{n-4}.
\label{METN4}
\end{eqnarray} 
Again we use the minimalist approach which allows us to write
\begin{eqnarray}
C_0^{\mu\nu\ss\tau}&=&\kk_0(A^{\mu\nu}A^{\ss\tau}+\Ab^{\mu\nu}\Ab^{\ss\tau}),\nonumber\\
C^{\mu\nu\ss\tau}&=&\kk(A^{\mu\nu}A^{\ss\tau}+\Ab^{\mu\nu}\Ab^{\ss\tau}). 
\end{eqnarray}
From eq(\ref{VACP11}) we find
\be
\kk_0=\kk\left(1-\fr{e^2}{6\pi^2}\fr{1}{n-4}\right)+\fr{e^2}{6\pi^2}\fr{c^2}{n-4}.
\label{METN5}
\ee

In order to discuss the renormalisation of the electron parameters it is necessary
to consider the lowest order photon propagator. Although the representation for the photon propagator 
in eq(\ref{INVREP}) is useful for exhibiting the pole divergences at $n=4$ in a general context,
it is implicitly a power series in the WLT associated with birefringence and the breakdown of Lorentz
invariance. In the present case of Petrov class N, it is possible and more convenient to obtain a
complete expression for the photon propagator that can be used in perturbation theory
calculations. We will choose the gauge so that $\LL^{\mu\nu}=g^{\mu\nu}$.
In lowest order in $e^2$ the inverse photon propagator is given by
\be
M^{\mu\ss}(q)=\Mc^{\mu\ss}(q)=q^2g^{\mu\nu}-C^{\mu\nu\ss\tau} q_\nu q_\tau.
\ee
That is, for Petrov class N,
\be
M^{\mu\ss}(q)=q^2g^{\mu\ss}-\kk(P^\mu P^\ss+\Pb^\mu\Pb^\ss),
\ee
where 
\be
P^\mu=A^{\mu\nu}q_\nu=l^\mu m^\nu(s)q_\nu-m^\mu(s)l^\nu q_\nu=l^\mu m^\nu q_\nu-m^\mu l^\nu q_\nu,
\ee
and 
\be
\Pb^\mu=\Ab^{\mu\nu}q_\nu=l^\mu \mb^\nu(s) q_\nu-\mb^\mu(s)l^\nu q_\nu=l^\mu \mb^\nu q_\nu-\mb^\mu l^\nu q_\nu.
\ee
If we set $P_\mu=g_{\mu\nu}P^\nu$ and $\Pb_\mu= g_{\mu\nu}\Pb^\nu$ then
\be
P^2=P^\mu P_\mu=\Pb^2=\Pb^\mu\Pb_\mu=0,
\ee
and
\be
P.\Pb=P^\mu\Pb_\mu=-(l.q)^2=-(l^\mu q_\mu)^2.
\ee
It is then easily verified that the inverse of $M^{\mu\nu}(q)$ is
\be
M_{\mu\nu}(q)=\fr{1}{q^2}g_{\mu\nu}+\kk\fr{(P_\mu P_\nu+\Pb_\mu\Pb_\nu)}{(q^2-\kk(l.q)^2)(q^2+\kk(l.q)^2)}
                        -\kk^2\fr{(l.q)^2(P_\mu\Pb_\nu+\Pb_\mu P_\nu)}{q^2(q^2-\kk(l.q)^2)(q^2+\kk(l.q)^2)}.
\ee
This may be rewritten as
\be
M_{\mu\nu}(q)=\fr{1}{q^2}g_{\mu\nu}
+\kk\fr{(P_\mu P_\nu+\Pb_\mu\Pb_\nu)}{(g^{(-)\aa\bb}q_\aa q_\bb)(g^{(+)\aa\bb}q_\aa q_\bb)}
               -\kk^2\fr{(l.q)^2(P_\mu\Pb_\nu+\Pb_\mu P_\nu)}{q^2(g^{(-)\aa\bb}q_\aa q_\bb)(g^{(+)\aa\bb}q_\aa q_\bb)},
\ee
where
\be
g^{(\pm)\aa\bb}=g^{\aa\bb}\pm\kk l^\aa l^\bb.
\ee
In discussing the divergence structure of the electron propagator we find from eq(\ref{PROPREN1A}) that
\be
\SS(0)=\SS^{(1)}+\SS^{(2)}+\SS^{(3)},
\ee
where
\be
\SS^{(1)}=ie^2m\mu^{4-n}\int\fr{d^nq}{(2\pi)^n}\fr{\gb^{\mu\nu}g_{\mu\nu}}
       {q^2(\gb^{\aa\bb} q_\aa q_\bb-m^2)},
\ee
\be
\SS^{(2)}=ie^2m\mu^{4-n}\int\fr{d^nq}{(2\pi)^n}\fr{\gb^{\mu\nu}(P_\mu P_\nu+\Pb_\mu\Pb_\nu)}
     {(g^{(+)\aa\bb} q_\aa q_\bb)(g^{(-)\aa\bb} q_\aa q_\bb)(\gb^{\aa\bb}q_\aa q_\bb-m^2)},
\ee
\be
\SS^{(3)}=-ie^2m\mu^{4-n}\int\fr{d^nq}{(2\pi)^n}\fr{(l.q)^2\gb^{\mu\nu}(P_\mu\Pb_\nu+\Pb_\mu P_\nu)}
    {q^2(g^{(+)\aa\bb} q_\aa q_\bb)((g^{(-)\aa\bb}q_\aa q_\bb)g^{\aa\bb} q_\aa q_\bb)(\gb^{\aa\bb}q_\aa q_\bb-m^2)}.
\ee
It is easily checked that $\gb^{\mu\nu}g_{\mu\nu}=n$, $\gb^{\mu\nu}P_\mu P_\nu=\gb^{\mu\nu}\Pb_\mu\Pb_\nu=0$
and $\gb^{\mu\nu}P_\mu\Pb_\nu=-(l.q)^2$. We have then
\be
\SS^{(1)}=ie^2m\mu^{4-n}nI,
\ee
where
\be
I=\int\fr{d^nq}{(2\pi)^n}\fr{1}{q^2(\gb^{\aa\bb} q_\aa q_\bb-m^2)},
\ee
\be
\SS^{(2)}=0,
\ee
and
\be
\SS^{(3)}=2ie^2m\mu^{4-n}l^\mu l^\nu l^\ss l^\tau I_{\mu\nu\ss\tau},
\ee
where
\be
I_{\mu\nu\ss\tau}=\int\fr{d^nq}{(2\pi)^n}\fr{q_\mu q_\nu q_\ss q_\tau}
    {q^2(g^{(+)\aa\bb} q_\aa q_\bb)((g^{(-)\aa\bb}q_\aa q_\bb)g^{\aa\bb} q_\aa q_\bb)(\gb^{\aa\bb}q_\aa q_\bb-m^2)}.
\ee
In appendix (\ref{INTEGRALS}) we show that
\be
I\simeq -\fr{i}{8\pi^2}\fr{1}{n-4},
\ee
and, making use of results for integrals listed there we can show that
\be
l^\mu l^\nu l^\ss l^\tau I_{\mu\nu\ss\tau}=0.
\ee
It follows that
\be
\SS(0)\simeq\fr{e^2}{2\pi^2}m\fr{1}{n-4}.
\ee
We have also, referring to eq(\ref{PROPREN1D}) and eq(\ref{PROPREN1E}), 
\be
H^{(1)\tau}_{~~~~~\rho}=H^{(11)\tau}_{~~~~~~\rho}+H^{(12)\tau}_{~~~~~~\rho}+H^{(13)\tau}_{~~~~~~\rho},
\ee
and
\be
H^{(2)\tau}_{~~~~~\rho}=H^{(21)\tau}_{~~~~~~\rho}+H^{(22)\tau}_{~~~~~~\rho}+H^{(23)\tau}_{~~~~~~\rho},
\ee
where
\be
H^{(11)\tau}_{~~~~~~\rho}
     =ie^2\mu^{4-n}(\dd^\mu_\rho\gb^{\tau\nu}+\dd^\nu_\rho\gb^{\tau\mu}-\dd^\tau_\rho\gb^{\mu\nu})g_{\mu\nu}I,
\ee
\be
H^{(12)\tau}_{~~~~~~\rho}=ie^2\mu^{4-n}(\dd^\mu_\rho\gb^{\tau\nu}+\dd^\nu_\rho\gb^{\tau\mu}-\dd^\tau_\rho\gb^{\mu\nu})
             \kk K_{\mu\nu},
\ee
where
\be
K_{\mu\nu}=\int\fr{d^nq}{(2\pi)^n}\fr{P_\mu P_\nu+\Pb_\mu\Pb_\nu}
                       {(g^{(+)\aa\bb}q_\aa q_\bb)(g^{(-)\aa\bb}q_\aa q_\bb)(\gb^{\aa\bb}q_\aa q_\bb)-m^2)}.
\ee
\be
H^{(13)\tau}_{~~~~~~\rho}=-ie^2\mu^{4-n}(\dd^\mu_\rho\gb^{\tau\nu}+\dd^\nu_\rho\gb^{\tau\mu}-\dd^\tau_\rho\gb^{\mu\nu})
           \kk^2J_{\mu\nu}, 
\ee
with
\be
J_{\mu\nu}=\int\fr{d^nq}{(2\pi)^n}\fr{(l.q)^2(P_\mu\Pb_\nu+\Pb_\mu P_\nu)}
                   {q^2(g^{(+)\aa\bb}q_\aa q_\bb)(g^{(-)\aa\bb}q_\aa q_\bb)(\gb^{\aa\bb}q_\aa q\bb-m^2)}.
\ee
With the aid of integrals evaluated in appendix \ref{INTEGRALS} it can be shpown that
\be
K_{\mu\nu}\simeq 0.
\ee
and
\be
J_{\mu\nu}\simeq 0.
\ee
Hence
\be
H^{(1)\tau}_{~~~~~\rho}\simeq\fr{e^2}{8\pi^2}(\dd^\mu_\rho\gb^{\tau\nu}+\dd^\nu_\rho\gb^{\tau\mu}-\dd^\tau_\rho\gb^{\mu\nu})g_{\mu\nu}\fr{1}{n-4}.
\ee
We have
\be
H^{(21)\tau}_{~~~~~~\rho}=-2ie^2\mu^{4-n}\gb^{\tau\bb}I'_{\bb\ss}(\dd^\mu_\rho\gb^{\ss\nu}+\dd^\nu_\rho\gb^{\ss\mu}-\dd^\ss_\rho\gb^{\mu\nu})g_{\mu\nu},
\ee
where
\be
I'_{\bb\ss}=\int\fr{d^nq}{(2\pi)^n}\fr{q_\bb q_\ss}{q^2(\gb^{\aa\bb}q_\aa q_\bb-m^2)^2},
\ee
\be
H^{(22)\tau}_{~~~~~~\rho}=-2ie^2\kk\mu^{4-n}\gb^{\tau\bb}
                     (\dd^\mu_\rho\gb^{\ss\nu}+\dd^\nu_\rho\gb^{\ss\mu}-\dd^\ss_\rho\gb^{\mu\nu})K'_{\aa\bb\mu\nu},
\ee
where
\be
K'_{\bb\ss\mu\nu}=\int\fr{d^nq}{(2\pi)^n}\fr{q_\bb q_\ss(P_\mu P_\nu+\Pb_\mu\Pb_\nu)}
                          {(g^{(+)\aa\bb}q_\aa q_\bb)(g^{(-)\aa\bb}q_\aa q_\bb)(\gb^{\aa\bb}q_\aa q_\bb-m^2)^2},
\ee
\be
H^{(23)\tau}_{~~~~~~\rho}=2ie^2\kk^2\mu^{4-n}\gb^{\tau\bb}
      (\dd^\mu_\rho\gb^{\ss\nu}+\dd^\nu_\rho\gb^{\ss\mu}-\dd^\ss_\rho\gb^{\mu\nu})
                     l^\xi l^\eta K'_{\bb\ss\xi\eta\mu\nu},
\ee
where
\be
K'_{\bb\ss\xi\eta\mu\nu}=\int\fr{d^nq}{(2\pi)^n}\fr{q_\bb q_\ss q_\xi q_\eta(P_\mu\Pb_\nu+\Pb_\mu P_\nu)}
                          {q^2(g^{(+)\aa\bb}q_\aa q_\bb)(g^{(-)\aa\bb}q_\aa q_\bb)(\gb^{\aa\bb}q_\aa q_\bb-m^2)^2},.
\ee
Again using the integrals evaluated in appendix \ref{INTEGRALS} it can be shown that 
$H^{(22)\tau}_{~~~~~~\rho}\simeq H^{(23)\tau}_{~~~~~~\rho}\simeq 0$. Hence
\be
H^{(2)\tau}_{~~~~~~\rho}(0)\simeq-\fr{e^2}{4\pi^2}(\dd^\mu_\rho\gb^{\tau\nu}+\dd^\nu_\rho\gb^{\tau\mu}-\dd^\tau_\rho\gb^{\mu\nu})g_{\mu\nu}\fr{1}{n-4}.
\ee
It follows that
\be
H^\tau_{~~\rho}(0)=-\fr{e^2}{8\pi^2}(\dd^\mu_\rho\gb^{\tau\nu}+\dd^\nu_\rho\gb^{\tau\mu}-\dd^\tau_\rho\gb^{\mu\nu})g_{\mu\nu}\fr{1}{n-4},
\ee
and
\be
H=H^\tau_{~~\tau}(0)=-\fr{e^2}{2\pi^2}\fr{1}{n-4}.
\ee
We have then
\be
h^\tau_{~~\rho}=-\fr{e^2}{4\pi^2}(\gb^{\tau\mu}g_{\mu\rho}-\dd^\tau_\rho)\fr{1}{n-4}.
\ee
Introducing the expressions for $\gb^{\mu\nu}$ we find
\be
h^\tau_{~~\rho}=\fr{e^2}{6\pi^2}(cP^\tau_{~~\rho}(s)+(w+4sc)l^\tau l_\rho)\fr{1}{n-4}.
\ee
We find
\be
h^\mu_{~~\rho}\gb^{\rho\nu}=\fr{e^2}{6\pi^2}{cP^\mu\nu}+(w-2c^2))\fr{1}{n-4}.
\ee
Recall that $e^2g^{(1)\mu\nu}=-h^\mu_{~~\rho}\gb^{\rho\nu}-h^\nu_{~~\rho}\gb^{\rho\mu}$ we find
\begin{eqnarray}
u_0&=&u-\fr{e^2}{3\pi^2}\fr{c}{n-4}.\nonumber\\
v_0&=&v-\fr{e^2}{3\pi^2}\fr{w-2c^2}{n-4}.
\label{METN6}
\end{eqnarray}

\subsubsection{\label{RENGP_N} Renormalisation Group for Bimetric Theory - Petrov Class N}

From eqs(\ref{METN4}) using the independence of the bare charge on the renormalisation scale 
we find
\be
\fr{de^2}{dt}=e^2\left((n-4)+\fr{e^2}{6\pi^2}\right).
\ee
In 4 dimensions this becomes
\be
\fr{de^2}{dt}=\fr{(e^2)^2}{6\pi^2},
\ee
with the solution
\be
\fr{e^2}{e_S^2}=\left(1-\fr{e_S^2}{6\pi^2}t\right)^{-1},
\ee
where $e_S$ is the value of the coupling when $\mu=\mu_S$.

From eqs(\ref{METN4})and eqs(\ref{METN6})e have the result
\be
u_0-s_0=c\left(1-\fr{e^2}{2\pi^2}\fr{1}{n-4}\right).
\ee
Since $u_0$ and $s_0$ are independent of the renormalisation scale 
it follows that
\be
\fr{1}{c}\fr{dc}{dt}=\fr{e^2}{2\pi^2}
\ee
The solution is
\be
\fr{c}{c_S}=\left(1-\fr{e_S^2}{6\pi^2}t\right)^{-3},
\ee
$c_S$ being the value of the coupling $c$ at the scale $\mu_S$.
Similarly
\be
v_0-r_0=w-\fr{e^2}{2\pi^2}\fr{w-c^2}{n-4},
\ee
leading to 
\be
\fr{dw}{dt}=\fr{e^2}{2\pi^2}(w-c^2),
\ee
with the solution
\be
w=\left(w_S-\fr{c_S^2}{2\pi^2}\left(\left(1-\fr{e_S^2}{6\pi^2}t\right)^{-6}-1\right)\right)\left(1-\fr{e_S^2}{2\pi^2}t\right)^{-3}.
\ee

We see immediately that in the infra red limit $\mu\rightarrow 0$ or $t\rightarrow -\infty$
\begin{eqnarray}
e^2&\rightarrow 0\nonumber\\
c&\rightarrow 0\nonumber\\
w&\rightarrow 0
\end{eqnarray}
Hence as we expect the infrared limit is the weak coupling limit and in this limit
both $c$ and $w$ vanish bringing the metrics $g^{\mu\nu}$ and $\gb^{\mu\nu}$ into
coincidence thus potentially removing the breakdown of Lorentz invariance, at least in the massless case. 
We have also from eq(\ref{METN5}) 
\be
\fr{d\kk}{dt}=\fr{e^2}{6\pi^2}\kk-\fr{e^2}{6\pi^2}c^2.
\ee
Using the above results we have
\be
\kk=\left(\kk_S-\fr{c_S^2}{5}\left(\left(1-\fr{e_S^2}{6\pi^2}t\right)^{-5}-1\right)\right)\left(1-\fr{e_S^2}{6\pi^2}t\right)^{-1}.
\ee
It follows that in the infrared limit $\kk$ vanishes and hence any birefringence.

Note that it we can choose $c_S=\kk_S=0$ which implies $c=\kk=0$ and therefore we can consistently set 
$s=u=\kk=0$. This leaves $w$ as the only significant remaining variable which provides a
Lorentz breaking scenario that is the lightlike case of Petrov class O referred to in the previous section.
Were we then to choose $w_S=0$ we would return to a situation of Lorentz invariance. If however $c_S\ne 0$ 
then we induce nonvanishing values for $w$. For the massless case we can again argue that Lorentz invariance 
returns in the infrared limit. However if we examine the renormalisation group for the mas  we find 
\be
m=m_S\left(1-\fr{e_S^2}{6\pi^2}t\right)^{9/4}.
\ee
This is the same behaviour as the Petrov class 0 model in the neighbourhood of the IR fixed point.
However in this case it holds for finite values of the Lorentz breaking parameters. Again we 
require a closer examination of the infrared limit in this case to deal with a non-zero mass
for the electron.

\section{\label{CONC} Conclusions}

We have examined the breakdown of Lorentz invariance in QED 
through a premetric formulation of elctrodynamics parametrised by a
tensor $U^{\mu\nu\ss\tau}$ that has the same symmetry properties
as the Riemann tensor. However we showed that in fact there is a {\it preferred} 
metric $g^{\mu\nu}$ that allows us to decompose $U^{\mu\nu\ss\tau}$ in the form
$$
U^{\mu\nu\ss\tau}=g^{\mu\ss}g^{\nu\tau}-g^{\mu\tau}g^{\nu\ss}-C^{\mu\nu\ss\tau}.
$$
where $C^{\mu\nu\ss\tau}$ has the symmetry properties of the Weyl tensor.
We can therefore use the Petrov classification for $C^{\mu\nu\ss\tau}$ to
delineate all the possible forms of Lorentz symmetry breaking in electrodynamics
and ultimately in QED. Apart from the null case for which there is no Lorentz
symmetry breaking in electrodynamis (QED requires further analyis) all the other
canonical examples exhibit birefringence. We established the dispersion relations
for each Petrov class. In all cases this has the form of homogeneous quartic
constraint on the wave vector of the mode. In some cases this quartic has two
quadratic factors each corresponding to a particular polarisation. Each factor
yields a separate and distinct light cone. In other cases the quartic does not factorise 
in this way and hence is inherently more complex than a simple double lightcone structure. 

In examining the plane wave solutions of the general Lorentz symmetry breaking case  we
made use of the gauge condition on the elctromagnetic field $g^{\mu\nu}\d_\mu A_\nu(x)=0$.
However, motivated by the potential absence of Lorentz symmetry we also explored a more general
gauge condition $\LL^{\mu\nu}\d_\mu A_\nu(x)=0$. The choice of gauge condiditon does not
affect the physical solutions but it does affect the unphysical ones. In fact $\LL^{\mu\nu}$
determines the light cone for these unphysical modes and also for the associated ghost
modes. The latter do not play any role in electrodynamics or QED but in a non-abelian gauge theory
they will do. In fact it turns out that the more general gauge condition
comes into its own when we consider the renormalisation program for QED.

We examine the general structure of renormalised BIMQED to one loop order in
perturbation theory but without assuming that the Lorentz symmetry breaking
is itself small. The nature of this breaking is determined by the metric $\gb^{\mu\nu}$
governing the propagation of the electron field through the tensor 
$W^{\mu\nu\ss\tau}=\gb^{\mu\ss}\gb^{\nu\tau}-\gb^{\mu\tau}\gb^{\nu\ss}$ which appears as
a factor in the residue of the UV divergence of the vacuum polarisation diagram. 
The standard decomposition of this tensor leads to a traceless Weil-like tensor $V^{\mu\nu\ss\tau}$.
The Petrov class of this tensor can be used to determine the nature of the Lorentz symmetry breaking 
in the model. We give examples, though by no means a complete list, of how different choices 
for $\gb^{\mu\nu}$ lead to different Petrov classes for $V^{\mu\nu\ss\tau}$.

Finally we apply the renormalisation program in detail to the two simplest Petrov
classes of symmetry breaking, namely class O and class N. We derive the renormalisation
group flows in these cases and conclude that Lorentz symmetry breaking is suppressed in the 
infra-red limit at least in the massless case. The results are entirely consistent with previous 
analyses. In our case we are not restricted to small deviations and can show, to $O(e^2)$, that the result
holds however large the breaking at shorter distances. That is there appear to be no 
unexpected fixed points for nonvanishing Lorentz symmetry breaking. 

It is of course of great interest to examine the corresponding case of a non-abelian
gauge theory such as QCD where the weak coupling fixed point occurs in the ultraviolet
rather than the infrared limit. We will consider this case in a later paper.

\appendix

\section{\label{GFIX2}Gauge Fixed Action for the EM Field}

The partition function for the electromagnetic field is $Z$ given by
\be
Z=\int d[A]e^{iS_{(p)}[A]}.
\label{PARTF1}
\ee
However $S_{(p)}[A]$ is invariant under the gauge transformation $A_\mu(x)\rightarrow A^h_\mu(x)=A_\mu(x)+\d_\mu h(x)$.
The expression for $Z$ contains therefore a factor $\int d[h]$ which we wish to extract. In anticipation of the 
gauge condition we wish to invoke namely
\be
\LL^{\mu\nu}\d_\mu A_\nu(x)=0,
\ee
we use the $\dd$-functional identity
\be
\int d[h]\DD\dd[C(x)-\LL^{\mu\nu}\d_\mu A^h_\nu(x)]=1,
\ee
where 
\be
\DD=\det \fr{\dd}{\dd h(x')}(\LL^{\mu\nu}\d_\mu A^h_\nu(x))=\det\LL^{\mu\nu}\d_\mu\d_\nu\dd(x-x').
\ee
We now rewrite  $Z$ as
\be
Z=\int d[h]\int d[A]\DD\dd[C(x)-\LL^{\mu\nu}\d_\mu A^h_\nu(x)]e^{iS_{(p)}[A]}.
\ee
Exploiting the gauge invariance of $S_{(p)}[A]$ and the measure $d[A]$, we can write this in the form
\be
Z=\int d[h]\int d[A]\DD\dd[C(x)-\LL^{\mu\nu}\d_\mu A_\nu(x)]e^{iS_{(p)}[A]}.
\label{GENF5}
\ee
Because it is a constant we can now drop the factor $\int d[h]$ in the above expression for $Z$. We could also drop the factor $\DD$
since for QED, it does not depend on the fields in the integrand. However we retain it in order to
elucidate the BRST transormation and to anticipate the corresponding results for non-Abelian gauge theories.
To this end we introduce anti-commuting ghost fields $c(x)$ and $\cb(x)$ and express $\DD$ (up to an irrelevant
constant factor) in the form
\begin{eqnarray}
\DD&=&\int d[c]d[\cb]\exp\{i\int d^nx\Om \cb(x)\LL^{\mu\nu}\d_\mu\d_\nu c(x)\},\nonumber\\
   &=&\int d[c]d[\cb]\exp\{-i\int d^nx\Om\d_\mu\cb(x)\LL^{\mu\nu}\d_\nu c(x)\}.
\end{eqnarray}
The second equality results from an integration by parts in the exponent.
Using arguments we are also free to replace $Z$ in eq(\ref{GENF5}) by
\begin{eqnarray}
Z&=&\int d[C]\exp\{-\fr{i}{2}\int d^nxC(x)^2\}\int d[A]\DD\dd[C(x)-\LL^{\mu\nu}\d_\mu A_\nu(x)]e^{iS_{(p)}[A]}
                                                                  \nonumber\\
   &=&\int d[A]d[c]d[\cb]\exp\{iS_{\mbox{g.f.}}\},
\label{GENF6}
\end{eqnarray}
where $S_{\mbox{g.f.}}$ is the full gauge fixed action for the photon sector given by
\be
S_{\mbox{g.f.}}=\int d^nx\{-\fr{1}{4}g^{\mu\ss}g^{\nu\tau}F_{\mu\nu}F_{\ss\tau}
                                            -\fr{1}{2}\LL^{\mu\nu}\LL^{\ss\tau}\d_\mu A_\nu(x)\d_\ss A_\tau(x)
                                             -\d_\mu\cb(x)\LL^{\mu\nu}\d_\nu c(x)\}.
\ee

The equation of motion for the photon field is
\be
g^{\mu\ss}g^{\nu\tau}\d_\mu\d_\ss A_\nu-(g^{\mu\ss}g^{\nu\tau}-\LL^{\mu\ss}\LL^{\nu\tau})\d_\nu\d_\ss A_\mu=0,
\ee
and those for the ghost fields are
\begin{eqnarray}
\LL^{\mu\nu}\d_\mu\d_\nu c(x)&=&0,\nonumber\\
\LL^{\mu\nu}\d_\mu\d_\nu\cb(x)&=&0.
\end{eqnarray}
Clearly $\LL^{\mu\nu}$ does play the role of the (inverse) metric for the ghost fields.

\section{\label{FALSING} Removal of Spurious Singularity}

We can verify the absence of the apparent singularity at $q^2=0$ 
by first introducing $\qh$ as the parity reflection of $q$. Of course $\qh^2=q^2$
and $q.\qh>0$ for $q\ne 0$. We then define the matrix $\Mch^{\nu\tau}$ given by
\be
\Mch^{\nu\tau}=\left(\dd^\nu_\rho+\fr{\qh^\nu q_\rho}{q^2}\right)\Mc^{\rho\ll}
                                       \left(\dd^\tau_\ll+\fr{q_\ll\qh^\tau}{q^2}\right).
\ee
That is
\be
\Mch^{\nu\tau}=\Mc^{\nu\tau}+q^\nu \qh^\tau+\qh^\nu q^\tau+\qh^\nu\qh^\tau.
\label{WF3}
\ee
It follows that $\det\Mch^{\nu\tau}$ is finite and in general nonvanishing on $q^2=0$.
From eq(\ref{WF3}) we find
\be
\Mc^{\rho\ll}=\left(\dd^\rho_\nu-\fr{\qh^\rho q_\nu}{q^2+\qh.q}\right)
                      \Mch^{\nu\tau}\left(\dd^\ll_\tau-\fr{q_\tau\qh^\ll}{q^2+q.\qh}\right).
\label{WF4}
\ee
From eq(\ref{WF4}) we obtain
\be
\Mc_{\rho\ll}=\left(\dd^\nu_\rho+\fr{q_\rho \qh^\nu}{q^2}\right)\Mch_{\nu\tau}
                  \left(\dd^\tau_\ll+\fr{\qh^\tau q_\ll}{q^2}\right),
\label{WF5}
\ee
where $\Mch_{\nu\tau}$ is the inverse of $\Mch^{\nu\tau}$. It is then straightforward
to express eq(\ref{WF2}) in the form
\be
M_{\nu\tau}=\left(\dd^\rho_\nu-\fr{q_\nu Q^\rho}{Q.q}\right)\Mch_{\rho\ll}
                 \left(\dd^\ll_\tau-\fr{Q^\ll q_\tau}{Q.q}\right)+\fr{q_\nu q_\tau}{(Q.q)^2}.
\label{WF6}
\ee
Eq(\ref{WF6}) shows clearly that $M_{\nu\tau}$ is singular only on the ghost mass-shell
and the surface yielding the dipersion relations for the physical states.

\section{\label{INTEGRALS} Special Integrals}

We will need the following integrals
\begin{eqnarray}
I&=&\int\fr{d^nq}{(2\pi)^n}\fr{1}{q^2(\gb^{\aa\bb} q_\aa q_\bb-m^2)}.\\
I'_{\mu\nu}&=&\int\fr{d^nq}{(2\pi)^n}\fr{q_\mu q_\nu}{q^2(\gb^{\aa\bb} q_\aa q_\bb-m^2)^2}.\\
I_{\mu\nu}&=&\int\fr{d^nq}{(2\pi)^n}\fr{q_\mu q_\nu}
         {(g^{(+)\aa\bb}q_\aa q_\bb)(g^{(-)\aa\bb}q_\aa q_\bb)(\gb^{\aa\bb} q_\aa q_\bb-m^2)}.\\
I'_{\mu\nu\ss\tau}&=&\int\fr{d^nq}{(2\pi)^n}\fr{q_\mu q_\nu q_\ss q_\tau}
         {(g^{(+)\aa\bb}q_\aa q_\bb)(g^{(-)\aa\bb}q_\aa q_\bb)(\gb^{\aa\bb} q_\aa q_\bb-m^2)^2}.\\
I_{\mu\nu\ss\tau}&=&\int\fr{d^nq}{(2\pi)^n}\fr{q_\mu q_\nu q_\ss q_\tau}
         {q^2(g^{(+)\aa\bb}q_\aa q_\bb)(g^{(-)\aa\bb}q_\aa q_\bb)(\gb^{\aa\bb} q_\aa q_\bb-m^2)}.\\
I'_{\mu\nu\ss\tau\xi\eta}&=&\int\fr{d^nq}{(2\pi)^n}\fr{q_\mu q_\nu q_\ss q_\tau q_\xi q_\eta}
         {q^2(g^{(+)\aa\bb}q_\aa q_\bb)(g^{(-)\aa\bb}q_\aa q_\bb)(\gb^{\aa\bb} q_\aa q_\bb-m^2)^2}.
\end{eqnarray}
\subsection{$I$}

The first integral we study is $I$. It can be put in the form
\be
I=(-i)^2\int_0^\infty du\int_0^\infty dv\int\fr{d^nq}{(2\pi)^n}\exp\{iu(g^{\aa\bb}q_\aa q_\bb+i\veps)
                                           +iv(\gb^{\aa\bb}q_\aa q_\bb-m^2+i\veps)\}.
\ee
If we change variables so that $u=x\ll$ and $v=(1-x)\ll$ we find
\be
I=(-i)^2\int_0^1dx\int_0^\infty d\ll\ll\int\fr{d^nq}{(2\pi)^n}\exp\{\ght^{\aa\bb}q_\aa q_\bb-(1-x)m^2+i\veps\},
\ee
where
\be
\ght^{\aa\bb}=xg^{\aa\bb}+(1-x)\gb^{\aa\bb}.
\ee
That is
\be
\ght^{\aa\bb}=g^{\aa\bb}+c(1-x)P^{\aa\bb}(s)+\aa(1-x)l^\aa l^\bb.
\ee
It follows that $\det \ght^{\aa\bb}=-1$. In this case the interpolating metric $\ght^{\aa\bb}$
never becomes singular. The constraints of causality as elucidated in ref \cite{} are automatically satisfied.
The same will be true of the generalised interpolating matrices encountered below.
On performing the $d^nq$ integral we find
\be
I=\fr{i}{(4\pi)^{n/2}}\int_0^1dx\int_0^\infty d(i\ll)(i\ll)^{1-n/2}\exp\{-i\ll[(1-x)m^2-i\veps]\}.
\ee
That is
\be
I =\fr{i}{(4\pi)^{n/2}}\int_0^1dx\GG(2-n/2)[(1-x)m^2-i\veps]^{n/2-2}. 
\ee
The pole at $n=4$ is then
\be
I\simeq -\fr{i}{8\pi^2}\fr{1}{n-4}.
\ee

\subsection{$I'_{\mu\nu}$}

Following the pattern of the previous calculation with appropriate changes to accommodate the changed powers
in the denominator we can put the integral in the form
\be
I'_{\mu\nu}=i\int dx(1-x)\int d\ll\ll^2\int\fr{d^nq}{(2\pi)^n}q_\mu q_\nu
                                                   \exp\{i\ll\ght^{\aa\bb}q_\aa q_\bb\}\exp\{-i\ll(1-x)m^2-i\veps\}.
\ee
Here $\ght^{\aa\bb}$ is the same as in the prvious example.
The invariance properties of the integral allow us to make the replacement
$$
q_\mu q_\nu\exp\{i\ll\ght^{\aa\bb}q_\aa q_\bb\}\rightarrow\fr{1}{n}\ght_{\mu\nu}(\ght^{\aa\bb}q_\aa q_\bb)
                                                           \exp\{i\ll\ght^{\aa\bb}q_\aa q_\bb\}
                                          =\fr{1}{n}\ght_{\mu\nu}\fr{\d}{\d(i\ll)}\exp\{i\ll\ght^{\aa\bb}q_\aa q_\bb\}
$$
We then have after performing the $d^nq$ and $d\ll$ integrations
\be
I'_{\mu\nu}=\fr{i}{2(4\pi)^{n/2}}\int dx(1-x)\ght_{\mu\nu}\GG(2-n/2)[(1-x)m^2-i\veps]^{n/2-2},
\ee
where
\be
\ght_{\mu\nu}=g_{\mu\nu}-c(1-x)P_{\mu\nu}(s)-((1-x)\aa+2(1-x)^2c^2)l_\mu l_\nu.
\ee
We have then
\be
I'_{\mu\nu}\simeq -\fr{i}{16\pi^2}\int dx(1-x)\ght_{\mu\nu}\fr{1}{n-4}.
\ee
Substituting the expression for $\ght_{\mu\nu}$ we obtain finally
\be
I'_{\mu\nu}=-\fr{i}{32\pi^2}\left\{g_{\mu\nu}-\fr{2}{3}cP_{\mu\nu}(s)-\left(\fr{2}{3}\aa+c^2\right)l_\mu l_\nu\right\}.
\ee

\subsection{$I_{\mu\nu}$}

We can express each propagator as before and obtain the representation
\be
I_{\mu\nu}=(-i)^3\int dudvdw\int\fr{d^nq}{(2\pi)^n}q_\mu q_\nu\exp\{u(g^{(+)\aa\bb}q_\aa q_\bb)+v(g^{(-)\aa\bb}q_\aa q_\bb)+w(\gb^{\aa\bb}q_\aa q_\bb-m^2)\}.
\ee
Using the transformation of integration variables $u=\ll x$, $v=\ll y$, $w=\ll z$ we obtain
\be
I_{\mu\nu}=i\int dxdydz\dd(1-x-y-z)\int d\ll\ll^2\int\fr{d^nq}{(2\pi)^n}q_\mu q_\nu\exp\{i\ght^{\aa\bb}q_\aa q_\bb\}\exp\{-i\ll[zm^2-i\veps]\},
\ee
where now
\be
\ght^{\aa\bb}=g^{\aa\bb}+zcP^{\aa\bb}(s)+(\kk(x-y)+\aa z)l^\aa l^\bb.
\ee
We have again $\det \ght^{\aa\bb}=-1$ and the inverse matrix is
\be
\ght_{\aa\bb}=g_{\aa\bb}-zcP_{\aa\bb}(s)-(\kk(x-y)+\aa z+2c^2z^2)l_\aa l_\bb.
\ee
As in the previous calculation we we make the substitution $q_\mu q_\nu\rightarrow \ght_{\mu\nu}(\ght^{\aa\bb}q_\aa q_\bb)/n$
and obtain after performing the $d^na$ and $d\ll$ integrations
\be
I_{\mu\nu}\simeq -\fr{i}{(4\pi)^2}\int dxdydz\dd(1-x-y-z)\ght_{\mu\nu} \fr{1}{n-4}.
\ee
Finally 
\be
I_{\mu\nu}=-\fr{i}{32\pi^2}\left\{g_{\mu\nu}-\fr{1}{3}cP_{\mu\nu}(s)-\fr{1}{3}(\aa+c^2)l_\mu l_\nu\right\}\fr{1}{n-4}.
\ee

\subsection{$I'_{\mu\nu\ss\tau}$}

With mild generalisations we can use the techniques of the previous calculations to obtain the result
\be
I'_{\mu\nu\ss\tau}\simeq-\fr{i}{32\pi^2}\int dxdydz\dd(1-x-y-z)z(\ght_{\mu\nu}\ght_{\ss\tau}+\cdots)\fr{1}{n-4},
\ee
where $\ght_{\mu\nu}$ is the same as in the previous calculation,
and $(\ght_{\mu\nu}\ght_{\ss\tau}+\cdots)=(\ght_{\mu\nu}\ght_{\ss\tau}+\ght_{\mu\ss}\ght_{\nu\tau}+\ght_{\mu\tau}\ght_{\ss\nu})$.

\subsection{$I_{\mu\nu\ss\tau}$} 

Again we can use the same style of calculation to obtain
\be
I_{\mu\nu\ss\tau}\simeq=-\fr{i}{32\pi^2}\int dxdydzdt\dd(1-x-y-z-t)(\ght_{\mu\nu}\ght_{\ss\tau}+\cdots)\fr{1}{n-4}.
\ee
Here
\be
\ght_{\mu\nu}=g_{\mu\nu}-2tc(+l_\mu(m_\nu(s)+\mb_\nu(s))+(m_mu(s)+\mb_\mu(s))l_\nu)-(\kk(y-z)-t\aa)l_\mu l_\nu.
\ee

\subsection{$I'_{\mu\nu\ss\tau\xi\eta}$}

Finally we obtain using the above style of calkculation
\be
I'_{\mu\nu\ss\tau\xi\eta}\simeq-\fr{i}{64\pi^2}\int dxdydzdt\dd(1-x-y-z-t)t(\ght_{\mu\nu}\ght_{\ss\tau}\ght_{\xi\eta}+\cdots)\fr{1}{n-4}.
\ee
Here $\ght_{\mu\nu}$ is the same as in the previous calculation and $(\ght_{\mu\nu}\ght_{\ss\tau}\ght_{\xi\eta}+\cdots)$
is a sum of 15 terms that symmetrise the exhibited term with respect to the labels $\{\mu\nu\ldots\eta\}$.

It is easy now to check that for each of the interpolating metrics above we have
\be
l^\aa\ght_{\aa\bb}=l_\bb,
\ee
and making use of this result we obtain the null results indicated in section \ref{RENEXN}.

\bibliography{mm2}
\bibliographystyle{unsrt}

\end{document}